\documentstyle{mn}
\input{epsfig.sty}

\newcommand{\Lya}{Ly$\alpha$~}
\newcommand{\A}{\AA~}
\newcommand{\apj}{ApJ}
\newcommand{\apjs}{ApJS}
\newcommand{\aj}{AJ}
\newcommand{\mnras}{MNRAS}

\newcommand{\araa}{ARA\&A}

\def\WH{W_{\rm H}}
\def\Gxz{\Gamma_{\rm HI}({\bf x},z)}
\def\GHI{\Gamma_{\rm HI}}
\newdimen\hssize
\newdimen\hdsize
\newdimen\hmsize
\begin{document}


\title[Ly$\alpha$ Leaks and Reionization]
      {Ly$\alpha$ Leaks and Reionization}

\author[Feng, Bi, Liu, Fang]
{Longlong Feng$^{1}$\thanks{E-mail: fengll@pmo.ac.cn}, Hongguang
Bi$^{1,2}$, Jiren Liu$^{2}$, Li-Zhi
Fang$^{2}$ \\
$^{1}$ Purple Mountain Observatory,Nanjing, 210008, China\\
$^{2}$ Department of Physics, University of Arizona, Tucson, AZ
85721, USA }


\date{}


\maketitle

\label{firstpage}


\begin{abstract}

Ly$\alpha$ absorption spectra of QSOs at redshifts $z\simeq6$ show
complete Gunn-Peterson absorption troughs (dark gaps) separated by
tiny leaks. The dark gaps are from the intergalactic medium (IGM)
where the density of neutral hydrogen are high enough to produce
almost saturated absorptions, however, where the transmitted leaks
come from is still unclear so far. We demonstrate that leaking can
originate from the lowest density voids in the IGM as well as the
ionized patches around ionizing sources using semi-analytical
simulations. If leaks were produced in lowest density voids, the IGM
might already be highly ionized, and the ionizing background should
be almost uniform; in contrast, if leaks come from ionized patches,
the neutral fraction of IGM would be still high, and the ionizing
background is significantly inhomogeneous. Therefore, the origin of
leaking is crucial to determining the epoch of
inhomogeneous-to-uniform transition of the the ionizing photon
background. We show that the origin could be studied with the
statistical features of leaks. Actually, Ly$\alpha$ leaks can be
well defined and described by the equivalent width $W$ and the full
width of half area $W_{\rm H}$, both of which are less contaminated
by instrumental resolution and noise. It is found that the
distribution of $W$ and $W_{\rm H}$ of Ly$\alpha$ leaks are
sensitive to the modeling of the ionizing background. We consider
four representative models: uniform ionizing background (model 0),
the photoionization rate of neutral hydrogen $\GHI$ and the density
of IGM are either linearly correlated (model I), or anti-correlated
(model II), and $\GHI$ is correlated with high density peaks
containing ionizing sources (model III). Although all of these
models can match to the mean and variance of the observed effective
optical depth of the IGM at $z\simeq 6$, their distribution of $W$
and $W_{\rm H}$ are very different from each other. Consequently,
the leak statistics provides an effective tool to probe the
evolutionary history of reionization at $z\simeq5-6.5$. Similar
statistics would also be applicable to the reionization of He II at
$z \simeq 3$

\end{abstract}




\section{Introduction}

In the last decade, the Ly$\alpha$ forests of QSO's absorption
spectra at redshifts $z \leq 5$ have played an important role in
understanding the diffuse cosmic baryon gas and the UV ionizing
photon background, and constraining cosmological models and
parameters (e.g. Rauch et al. 1997; Croft et al. 2002; Bolton et al.
2005; Seljak et al. 2005; Jena et al. 2005; Viel et al. 2006).
Recently, more and more UV photon sources, including QSOs, GRB,
Lyman-break galaxies, and Ly$\alpha$-emitters at redshifts $z>5$
have been observed (see Ellis 2007 and reference therein). Due to
the rapidly increase of Gunn-Peterson (GP) optical depth at $z > 5$,
their absorption spectra show long dark gaps on scales of tens of
Mpc separated by tiny transmitted leaks. It has been suggested that
we are observing the end stage of reionization (Fan et al. 2006).

It has been known that the dark gaps are from the IGM where the density
of neutral hydrogen are high enough to produce almost complete
absorptions, however, where the transmitted leaks come from is still unclear. In
photoionization equilibrium, the density of neutral hydrogen $n_{\rm
HI}\propto\alpha\rho^2$, here $\alpha$ is the recombination rate and
$\rho$ is the density of IGM; therefore, even when most of the IGM
are neutral enough to produce complete Ly$\alpha$ absorptions, it is
still possible for the lowest density voids to provide prominent
transmitted fluxes. On the other hand, the leaks can also come from
ionized patches around ionizing sources where the intensity of UV
radiation are higher than average.

The origin of leaking is crucial to understanding the history of
reionization. According to commonly accepted scenario of the
reionization, in the early stage, the ionized regions are isolated
patches in the neutral hydrogen background (e.g., Ciardi et al.
2003; Sokasian et al. 2003; Mellema et al. 2006; Gnedin 2006; Trac
\& Cen 2006). and the subsequent growing and overlapping of the
ionized patches lead to the ending of reionization (e.g., Ciardi et
al. 2003; Sokasian et al. 2003; Mellema et al. 2006; Gnedin 2004).
Thus, if leaks mostly come from ionized patches, reionization should
happen in the early stage. In contrast, if they were produced in lowest
density voids, the the UV ionizing background might has already
underwent an evolution from highly inhomogeneous to uniform
distribution.

A variety of statistics has been used to study the evolution of
reionization, such as the mean and dispersion of GP optical depth,
the probability distribution function (PDF) of the flux, and the
size of dark gaps (e.g., Fan et al. 2002, 2006; Songaila \& Cowie
2002; Paschos \& Norman 2005; Kohler et al. 2007; Gallerani et al.
2006; Becker et al. 2006), but all of them seem to be ineffective to
provide the information of leak's origin and the inhomogeneity of UV
ionizing background. The GP optical depth is an average, and not
sensitive to details of reionization. The statistical properties
(mean and variance) of the GP optical depth at $z\simeq 6$ can be
well explained by either the fluctuation of ionizing background (Fan
et al. 2006) or models with uniform ionizing background (Lidz et al.
2006; Liu et al. 2006, hereafter PaperI). The PDF of the flux is
also insensitive to the geometry of reionization. In addition, the
PDF is heavily contaminated by noise and distorted by resolution.

Dark gaps are defined to be continuous regions with optical depth
above a threshold in spectra. Intuitively, the statistics of dark
gap should contain the same information as leaks. However, the size
of dark gaps are sensitive to the instrumental resolution, because
higher resolution data contain more small leaks (e.g., Paschos \&
Norman 2005; PaperI), and they also are contaminated by
observational noise. Moreover, dark gaps are from saturated
absorptions, they are featureless and contains generally less information of
non-saturated  absorption.

In this paper we made a statistical approach to \Lya leaks. The
purpose is to show that the statistical features of \Lya leaks would
be effective tool to reveal the origin of \Lya leaks, and to probe
the evolution of reionization. Similar to \Lya absorption lines,
\Lya leaks have a rich set of statistical properties, such as the
width of leak profile. Unlike dark gaps, the properties of \Lya
leaks can be defined through integrated quantities, which are less
contaminated by resolution and noise. Moreover, \Lya leaks are from
regions of non-saturated absorptions and encode more information of
reionization; therefore, the \Lya statistics would provide more
underlying physics of reionization than all the above-mentioned
statistics.

The paper is organized as follows. \S 2 describes the method to
produce \Lya absorption samples. \S 3 presents the statistical
properties of Ly$\alpha$ leaks with a uniform ionizing background.
\S 4 analyzes the effect of inhomogeneous ionizing background.
Conclusion and discussion are given in \S 5.

\section{SIMULATION SAMPLES OF HIGH REDSHIFT Ly$\alpha$
   ABSORPTION SPECTRUM}

\subsection{Method}

We simulate Lyman series absorption spectra of QSOs between $z=3.5$
and 6.5 using the same lognormal method as those for low redshifts
$z \simeq 2 - 3$ (e.g., Bi et al. 1995; Bi \& Davidsen 1997). In
this model, the density field $\rho({\bf x})$ of the IGM is given by
an exponential mapping of the linear density field $\delta_0 ({\bf
x})$ as
\begin{equation}
\rho({\bf x}) = \bar{\rho}_0\exp[\delta_0 ({\bf x}) -  \sigma_0^2/2],
\end{equation}
where $\sigma^2_0=\langle \delta_0 ^2 \rangle $ is the variance of
the linear density field on scale of the Jeans length. Obviously,
the 1-point PDF of $\rho({\bf x})$ is lognormal. In this model, the
velocity field of baryon gas is produced by considering the
statistical relation between density and velocity field (Bi \&
Davidsen, 1997; Choudhury et al. 2001; Veil et al. 2002).

The dynamical bases of the lognormal model have gradually been
settled in recent years. First, although the evolution of cosmic
baryon fluid is governed by the Naiver-Stokes equation, the dynamics
of growth modes of the fluid can be sketched by a stochastic force
driven Burgers' equation (Berera \& Fang 1994). On the other hand,
the lognormal field is found to be a good approximation of the
solution of the Burgers' equation (Jones 1999). The one-point
distribution of the cosmic density and velocity fields on nonlinear
regime are consistent with lognormal distribution (e.g. Yang et al.
2001, Pando et al. 2002). Especially, it has been shown recently
that the velocity and density fields of the baryon matter of the
standard $\Lambda$CDM model is well described by the so-called
She-L\'ev\u{e}que's universal scaling formula, which is given by a
hierarchical process with log-Poisson probability distribution (He
et al. 2006, Liu \& Fang 2007).

The simulation is performed in the concordance $\Lambda$CDM cosmological model with
parameters  $\Omega_{m} = 0.27$, $h=0.71$, $\sigma_8 = 0.84$,
and $\Omega_{b} = 0.044$. The thermodynamic evolution in the IGM
is actually a rather complex process, because the nonlinear clustering leads to a
multi-phased IGM. As shown in cosmological hydrodynamic simulations (e.g., He et al. 2004),   
for a given mass density, the temperature of the IGM could have large scatters with differences 
up to two orders.  
Nevertheless, the equation of state in \Lya clouds can be well approximated by 
a polytropic relation with $\gamma =
4/3$ ( Hui \& Gnedin; He et al. 2004). The neutral fraction $f_{\rm
HI}$ is obtained by solving the photoionization equilibrium equation. The
photoionization rate $\Gamma_{\rm HI}$ will be given in \S 2.2. We
then construct synthetic absorption spectra by convoluting the neutral hydrogen 
density field with Voigt profiles. For each given $z$, the
size of the simulation samples is $\Delta z=0.3$, and there are
$2^{14}$ pixels in each simulation box.

\subsection{Redshift-Dependence of Photoionization Rate}

\begin{figure*}
\centerline{\psfig{figure=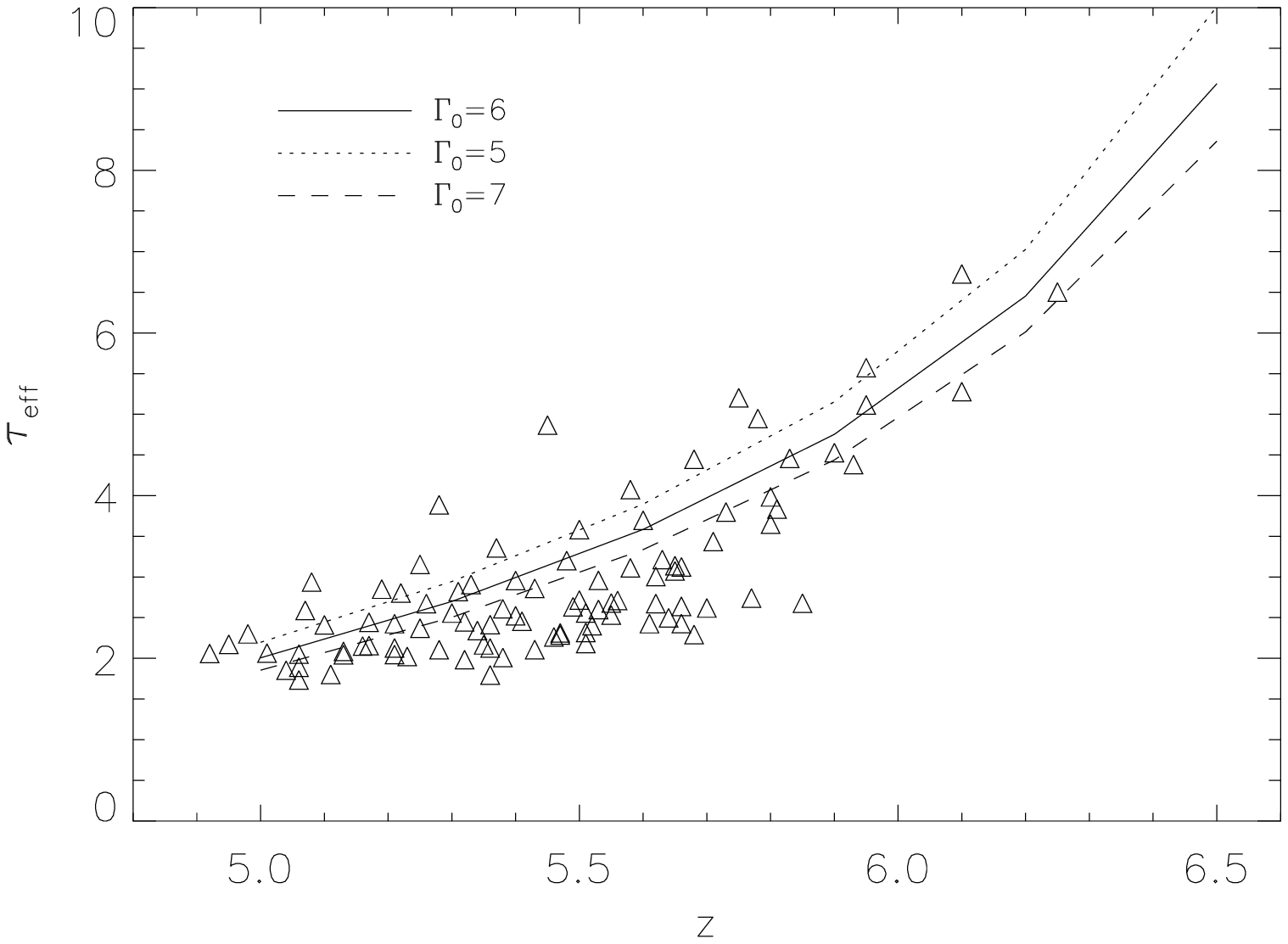,width=8.7truecm}
\psfig{figure=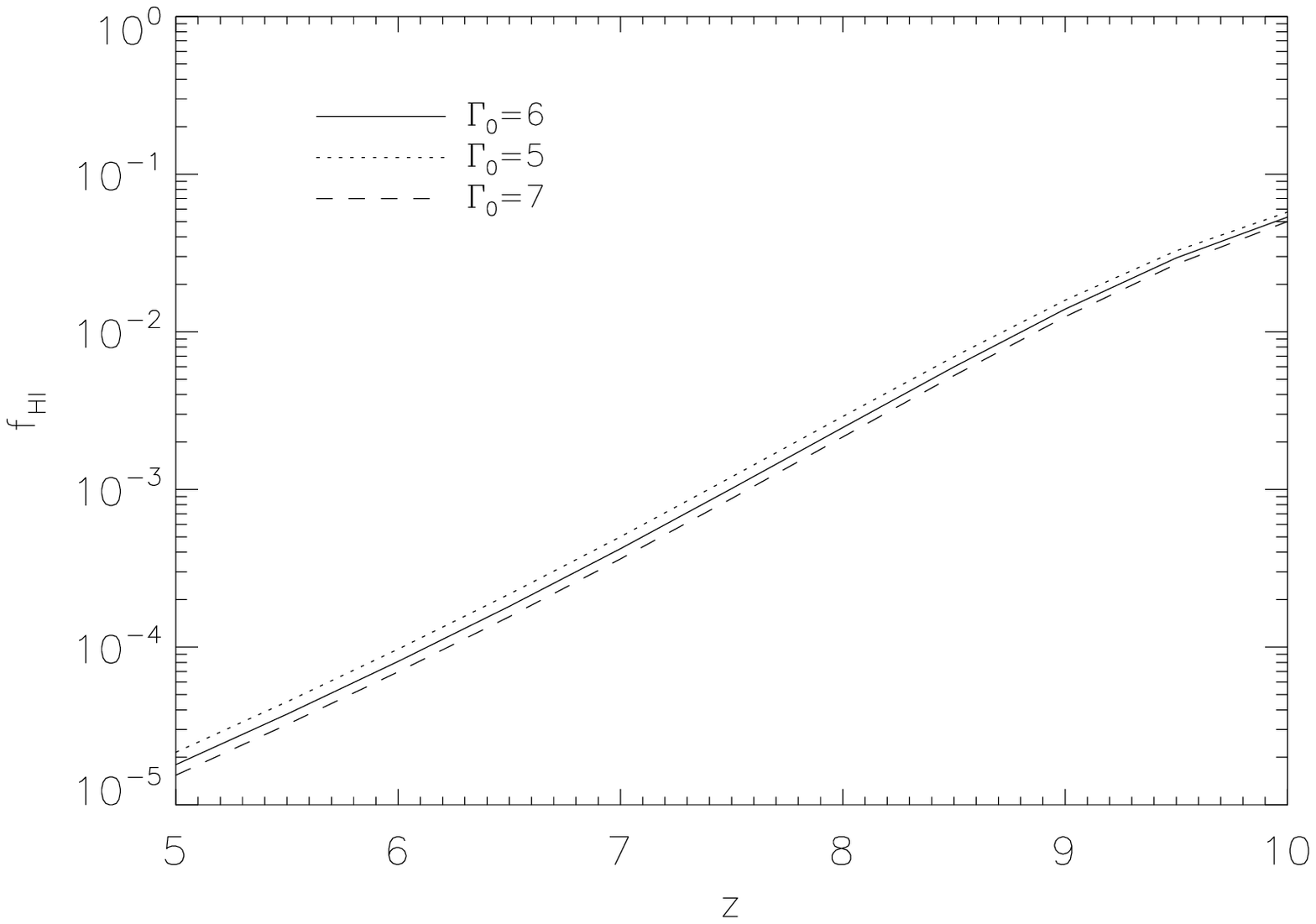,width=8.7truecm}}
\caption{Redshift evolution of effective optical depth
and neutral hydrogen fraction $f_{\rm HI}$ with the photoionization
rate given by eq.(2). The amplitude is taken to be $\Gamma_0=5$, 6
and 7. The data points are taken from the \Lya observation of Fan et
al. (2006).} \label{htb}
\end{figure*}

If the distribution of the IGM is uniform and the UV ionizing
background is independent of redshift, the mean GP optical depth of
\Lya absorption should approximately increase with redshift as
$(1+z)^{4.5}$. The observations of dark gaps directly show that the
GP optical depth undergoes a stronger evolution at $z \simeq 6$, and
consequently, the UV ionizing background would decrease rapidly with
redshift at $z\simeq 6$. The strong evolution scenario of the UV
ionizing background is supported by a number of simulations or
semi-analytical models of reionization (e.g., Razoumov et al. 2002;
Gnedin 2004; Oh \& Furlanetto 2005; Pascho \& Norman 2005; Wyithe \&
Loeb 2005; Kohler et al. 2007; Gallerani et al. 2006). It has been
found that an evolution of photoionization rate as follows can fit
the strong redshift evolution of the GP optical depth (Paper I) :
\begin{equation}
\Gamma_{\rm HI}(z) = \Gamma_0\exp\{-[(1+z)/(1+3.2)]^{2.4} \},
\end{equation}
which is in units of 10$^{-12}$ s$^{-1}$. Note that the power index
2.4 in equation (2) is little different from the one used in Paper I
because we use a different $T_0$ in this paper (also see below).

With eq.(2), we calculate the redshift dependencies of neutral
hydrogen fraction $f_{\rm HI}$ and the effective optical depth,
$\tau_{eff}\equiv -\ln(\overline{F})$, where $\overline{F}$ is the
mean of transmitted flux. The results are plotted in Figure 1. The
data points for $\tau_{eff}$ are taken from \Lya observations of Fan
et al. (2006). For best fitting, the amplitude $\Gamma_0$ is in the
range 5-7, which can be considered as the allowed range of
$\Gamma_{\rm HI}(z)$. In this paper, we will use $\Gamma_0=6$ as the
fiducial photoionization rate. It is interesting to note that
$f_{\rm HI}$ approaches to $\simeq 0.1$ at redshift $z\simeq 10$,
which is consistent with the electron scattering optical depth given
by the data of CMB polarization of WMAP III (Page et al. 2007).

It should be pointed out that the assumption of $T_0=2\times 10^{4}$
K (\S 2.1) is well reasonable at $z\leq 5$ (e.g. Hui \& Gnedin 1997;
He et al 2004) and may still be applicable at $z\simeq6$ if the mass
averaged neutral fraction of hydrogen is not larger than 10$^{-3}$,
and the photon heating rate is small. However, at higher redshift,
say $z \geq 6$, the temperature $T_0$ might be redshift-dependent.
Yet, no proper information on $T_0(z)$ is available at high
redshift, and this leads to uncertainty of the model. Fortunately,
in photoionization equilibrium, the neutral fraction $f_{\rm HI}$
depends mainly on a degenerate factor $\Gamma_{\rm
HI}(z)T^{0.75}_0(z)$. Thus, the problem with the uncertainty of
$T_0(z)$ can be overcame if we use the combined parameter of
$\Gamma_{\rm HI}(z)[T_0(z)/2\times 10^4]^{0.75}$ to fit the data. In
the range $z\leq 6$, this parameter actually is $\Gamma_{\rm
HI}(z)$; in the range of $z>6$, it is different from $\Gamma_{\rm
HI}(z)$ by a factor of $[T_0(z)/2\times 10^4]^{0.75}$. Thus, the
redshift-evolution of $\Gamma_{\rm HI}(z)$ would be slower than
eq.(2) if $T_0(z)$ is less than $2\times 10^4$ K at higher
redshifts.

\begin{figure*}
\centerline{\psfig{figure=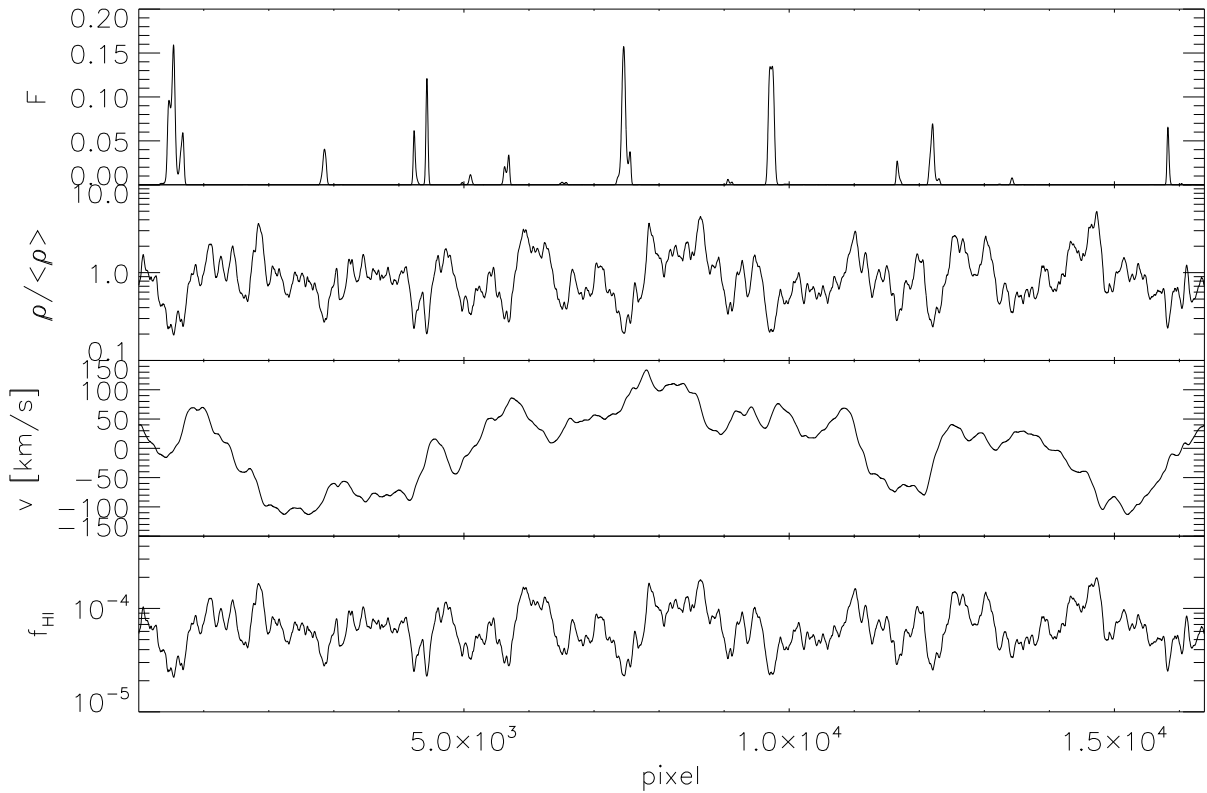,width=\hdsize}} \caption{An
example of simulated spectrum at $z=6$. It shows the transmitted
flux $F$, the density $\rho$ of baryon gas, the velocity $v$, and
the fraction of neutral hydrogen $f_{\rm HI}$ from top to bottom.
$\langle \rho\rangle$ is the mean of baryon matter.}
\end{figure*}

\subsection{An Example of Ly$\alpha$ Absorption Spectrum at $z=6$}

As an example of Ly$\alpha$ leaks, we plot a simulated sample of
\Lya absorption spectrum at $z=6$ with a uniform ionizing background
in Figure 2, which shows the transmitted flux $F$, the density
$\rho$ of baryon gas, the bulk velocity $v$, and the neutral
hydrogen fraction $f_{\rm HI}$. As expected, the mean of transmitted
flux is very small, about 0.004, and corresponds to an effective
optical depth 5.5. Nevertheless, we see spiky features with the
transmitted flux $F$ as large as $0.15$. They are leaks.

At low redshifts $z<5$, the \Lya forests in QSO's spectra have a
spectral filling factor significantly less than one and can be
decomposed into individual \Lya absorption lines. At redshifts
$z>5$, it is meaningless to decompose the spectra into individual
lines since almost the whole spectra are absorbed completely. We
note, however, the transmitted leaks look like emission features
upon the dark background, and the absorption spectra can be
decomposed into individual "emission lines", i.e., \Lya leaks.

Comparing the top, the second and bottom panels, we see that all the
leaks comes from the regions with mass density less than 0.3 of the
mean mass density of baryon gas. The neutral fraction for leaks is
$f_{\rm HI}\sim 2\times10^{-5}$, which yields a GP optical depth
$\sim2.5$ for overdensity 0.25 and a $F\sim0.1$, while the mean
neutral fraction of the entire example is about $7\times 10^{-5}$,
which is high enough to produce dark gaps. The column density of
neutral hydrogen of the leaking features is mainly in the range of
10$^{13}$-10$^{14}$ cm$^{-2}$, which are the non-saturated
absorption regions, and therefore, leaks can come from regions where
no enough neutral hydrogen to produce complete absorptions.

Similar to very high density clouds, $\rho/\bar{\rho}\gg 1$, the
regions with very low density $\rho/\bar{\rho}\ll 1$ are rare events
in the cosmic clustering. Therefore, leaks may provide valuable test
on models of clustering. For instance, the lognormal distributions
are long tailed in both high and low density sides, and it contains
more high density events as well as low density events than Gaussian
model. It should be emphasized that, once the photoionization rate
$\Gamma_{\rm HI}$ is determined from the GP optical depth, the
statistical property of the sample shown in Figure 2 doesn't contain
free parameters. These samples have been successfully used to
explain the following observations: 1.) the large dispersion of the
GP optical depth; 2.)the PDF of the flux, and 3.) the evolution of
the size of dark gaps (Paper I). Now we use them to study the
statistical properties of Ly$\alpha$ leaks.

\begin{figure}
\centerline{\psfig{figure=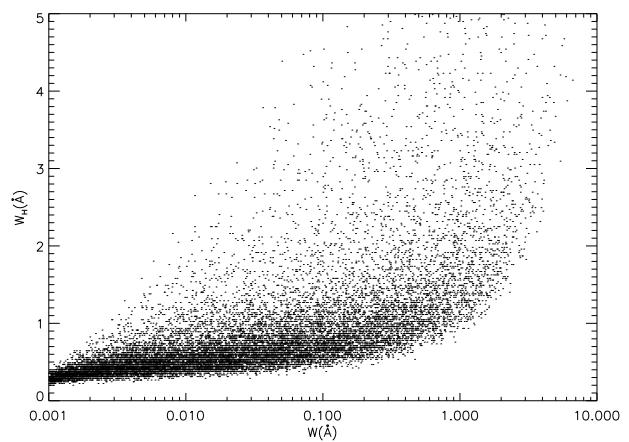,width=8.7truecm}}
\caption{Relation between $W$ and $W_{\rm H}$ of \Lya leaks at
$z=6$.}\label{HW}
\end{figure}

\section{STATISTICAL PROPERTIES OF Ly$\alpha$ LEAKS}

The \Lya absorption lines at low redshifts are described by well
defined quantities, such as the equivalent width, FWHM (full width
half maximum), and the Voigt profile, all of which are easily
related to physical interpretations. Since the transmitted leaks
look like emission features, one can decompose the spectra into
individual "emission lines" and describes it by quantities similar
to the \Lya absorption lines. In this section we study the
statistical distribution of \Lya leaks in the model with a uniform
ionizing background. The effects of inhomogeneous ionizing
background will be discussed in next section.

\subsection{Profile of Leaks}

\begin{figure*}
\centerline{\psfig{figure=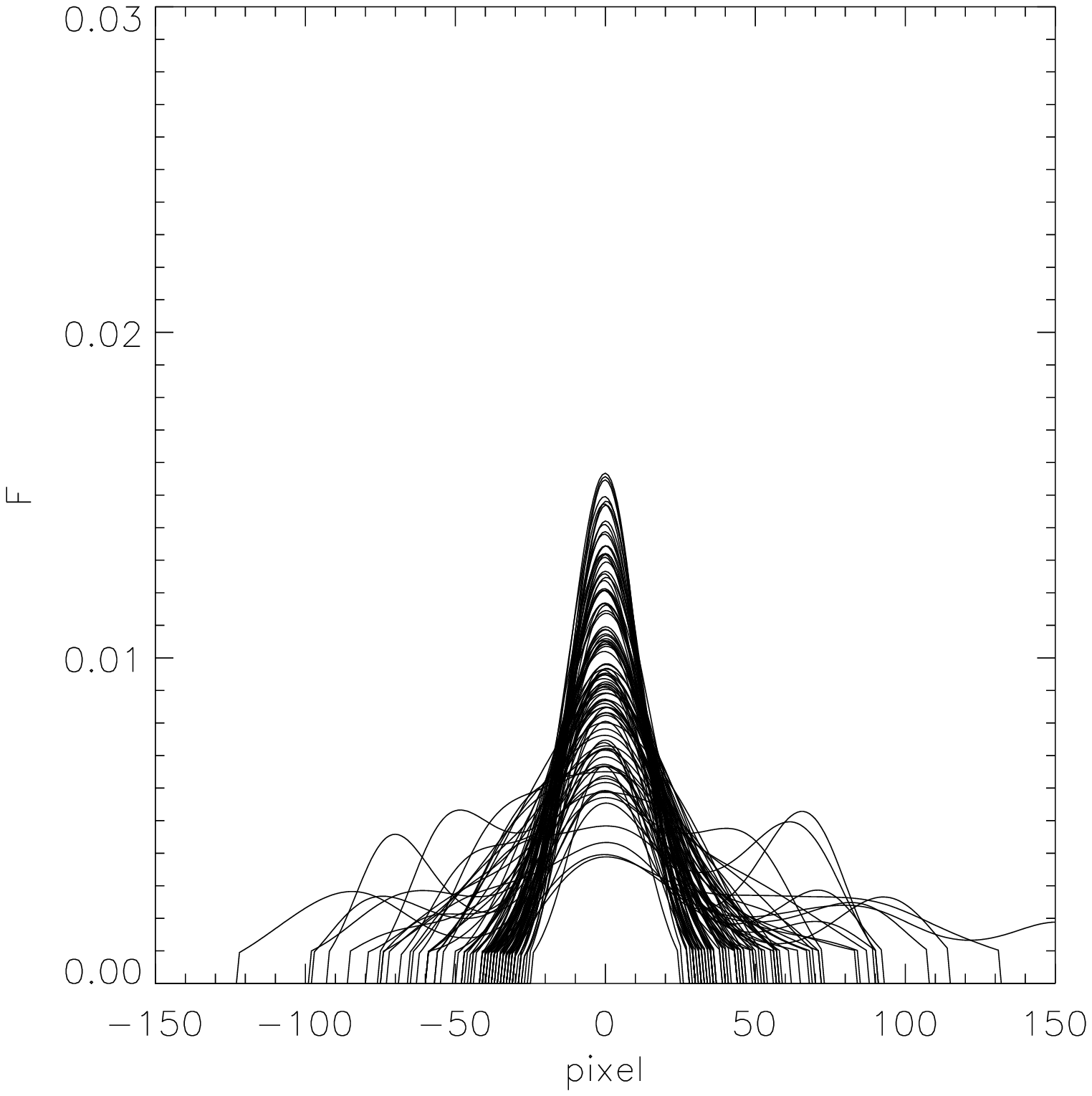,width=\hmsize}
\psfig{figure=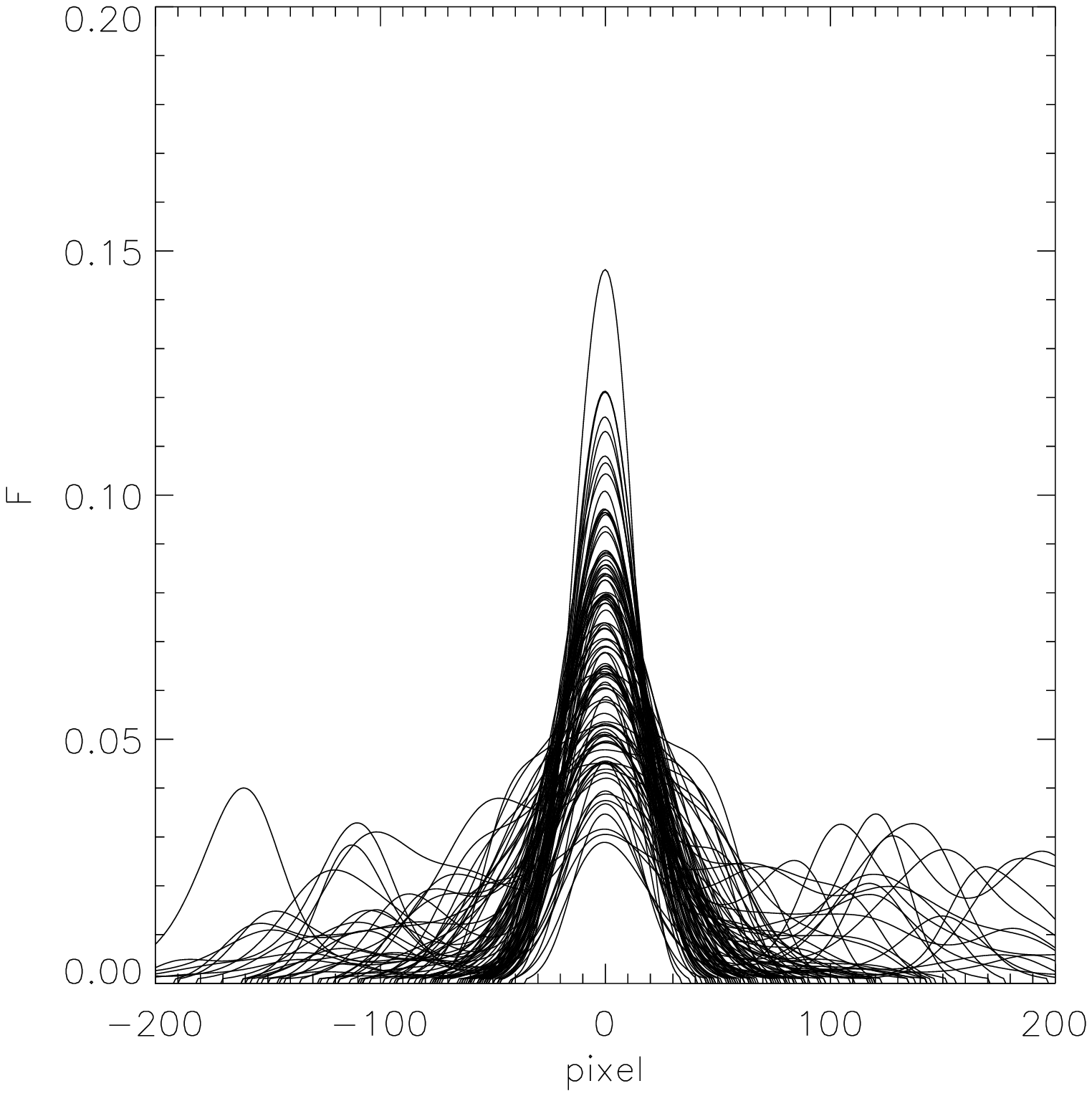,width=\hmsize}
\psfig{figure=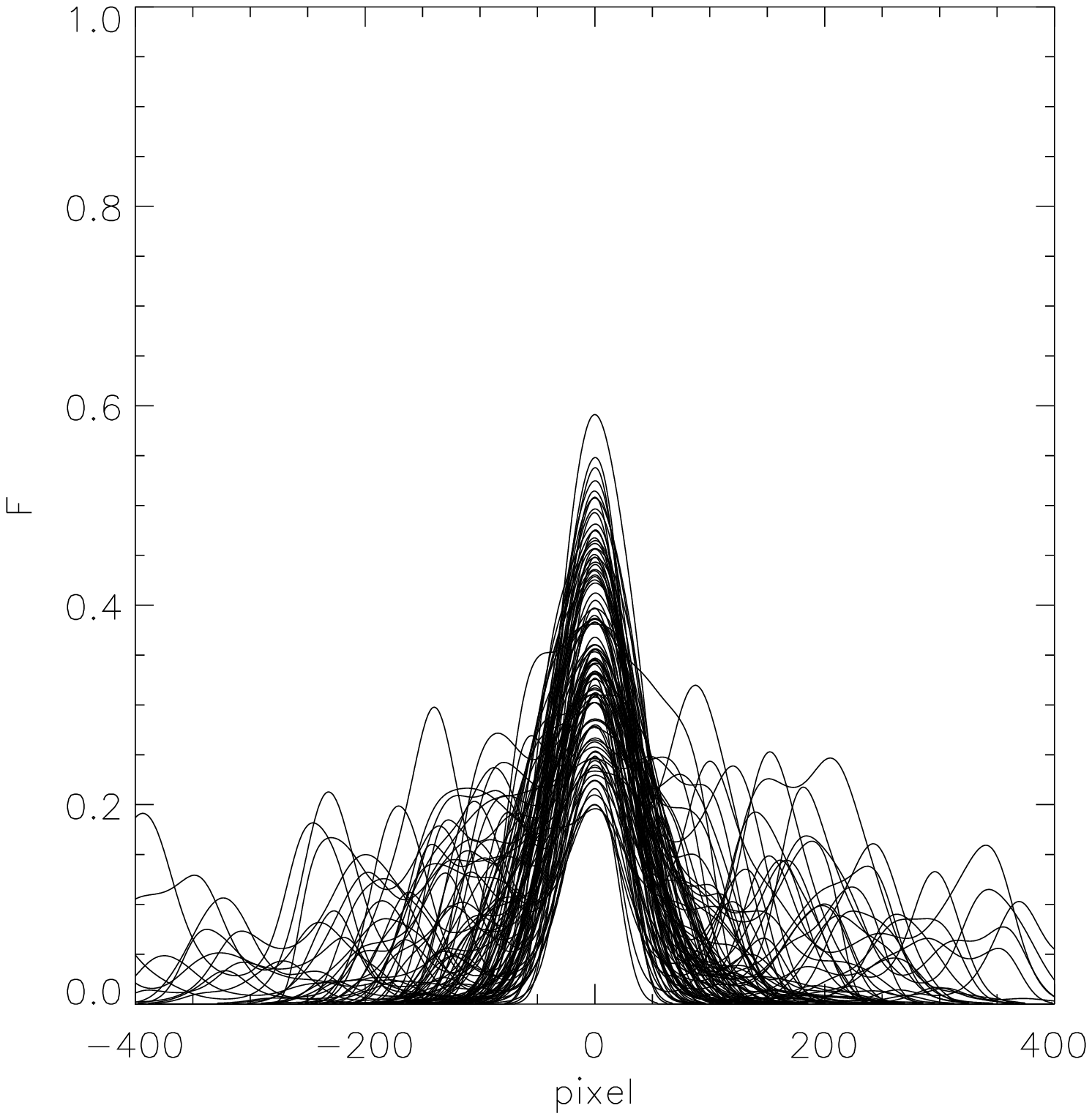,width=\hmsize}} \caption{Profiles of leaks
with equivalent widths $W=$0.01 (left), 0.1 (middle) and 1\AA \
(right) of Ly$\alpha$ leaks at $z=6$. One pixel is about
0.022\A.}\label{profile}
\end{figure*}

The center of a leak is identified as the maximum of transmitted
flux, and the boundaries are two nearby positions around the center
where the flux falls down to zero or to noise level (we take it to
be $F = 0.001$ in this paper). The properties of leaks can be
measured by two quantities: an equivalent width $W$, which is the
total area under the profile of a leak, and a half width $W_{\rm
H}$, which is defined as the width around center within which the
area under the profile of the leak is equal to half of the total
area of the leak.

The equivalent width $W$ and the half width $W_{\rm H}$ are two
independent measurements of \Lya leaks. This point can be seen in
Figure 3, which gives the value of $W$ and $W_{\rm H}$ for each leak
at $z=6$. It does not indicate a significant correlation between $W$
and $W_{\rm H}$, especially in the region of $W>0.1$ \A and $W_{\rm
H}>0.1$ \AA. Figure 3 also shows that the distribution of $W_{\rm
H}$ has a lower limit $0.2$ \AA, which is due to the Jeans scale
used for smoothing the sample. On the other hand, the equivalent
width $W$ distributes in the range from $0.001$ \A to $5$ \AA.
Obviously, one is unable to introduce two independent quantities for
characterizing dark gaps.

The profiles of 100 randomly sampled leaks with $W$ = 0.01, 0.1, and
1 \AA~ are displayed in Fig.\ref{profile}. The tails of the profiles
in the three panels of Fig.\ref{profile} look like the Lorentz
profile, and of course, they do not have the meaning of the natural
width of an absorption line. The profiles of leaks for a given
equivalent width $W$ have very large dispersions; for example, for
$W$=0.1\AA, the flux covers a range from $F=0.03$ to 0.15, and
$W_{\rm H}$ can be 0.4 to 4 \AA. As the leaks are formed out of the
two neighboring complete absorption troughs, which depend on
inhomogeneities of density, velocity, and temperature fields, the
large scatter of the profiles is expected.

Figure 4 also shows that some leaks may have multiple local maximums
above the noise level. For clarity and easy-operating, we treat them
as one leak.  The current
observational resolution is of the order of 10km/s, which
corresponds to $\Delta\lambda \sim 0.28$\A in observer's frame, or
about 13 pixels of simulation. Therefore, the observed \Lya leaks
would have a resolvable width.

\subsection{Number Density Distributions of Ly$\alpha$ Leaks}
\begin{figure*}
\centerline{\psfig{figure=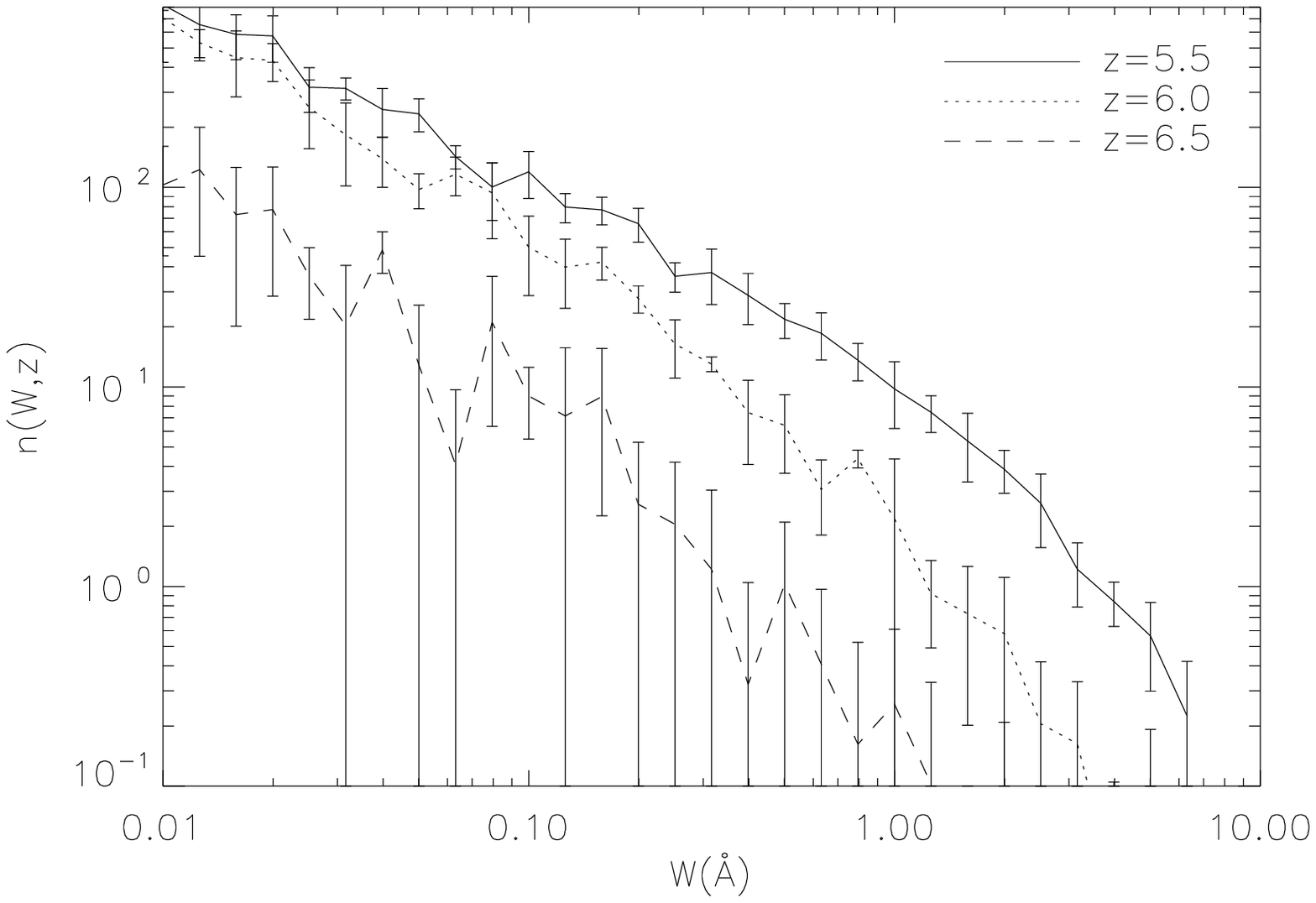,width=8.7truecm}
\psfig{figure=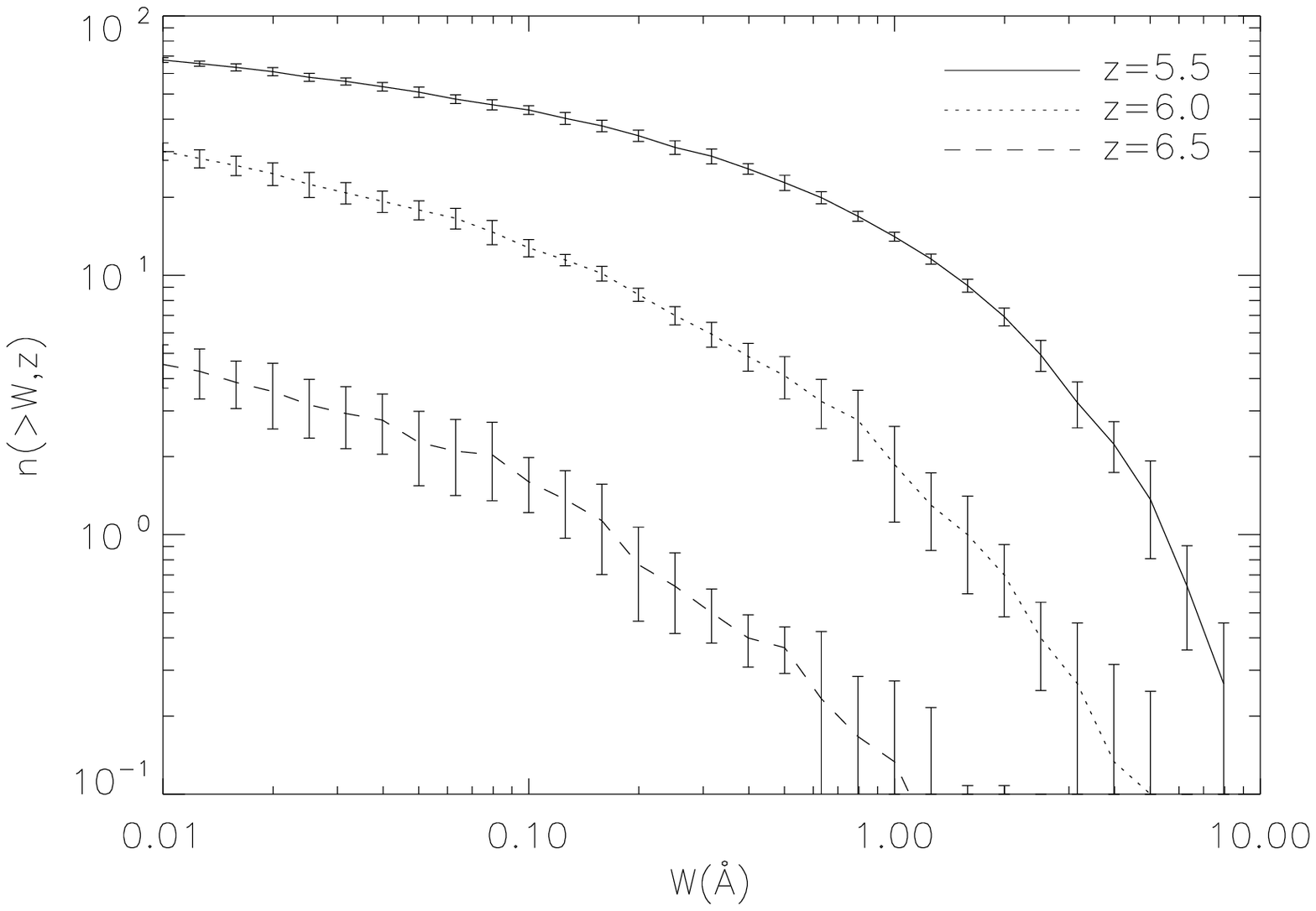,width=8.7truecm}}
\centerline{\psfig{figure=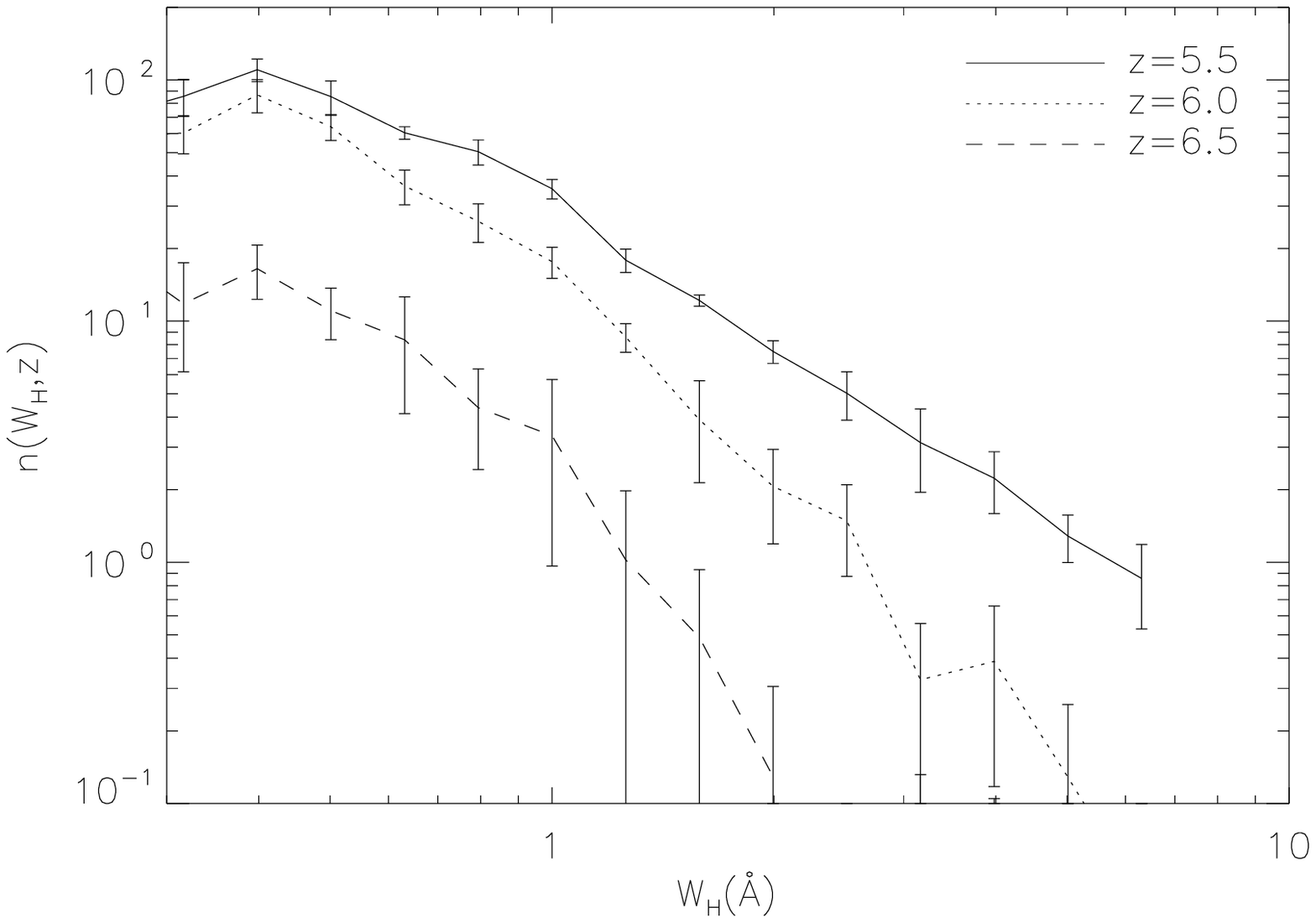,width=8.7truecm}
\psfig{figure=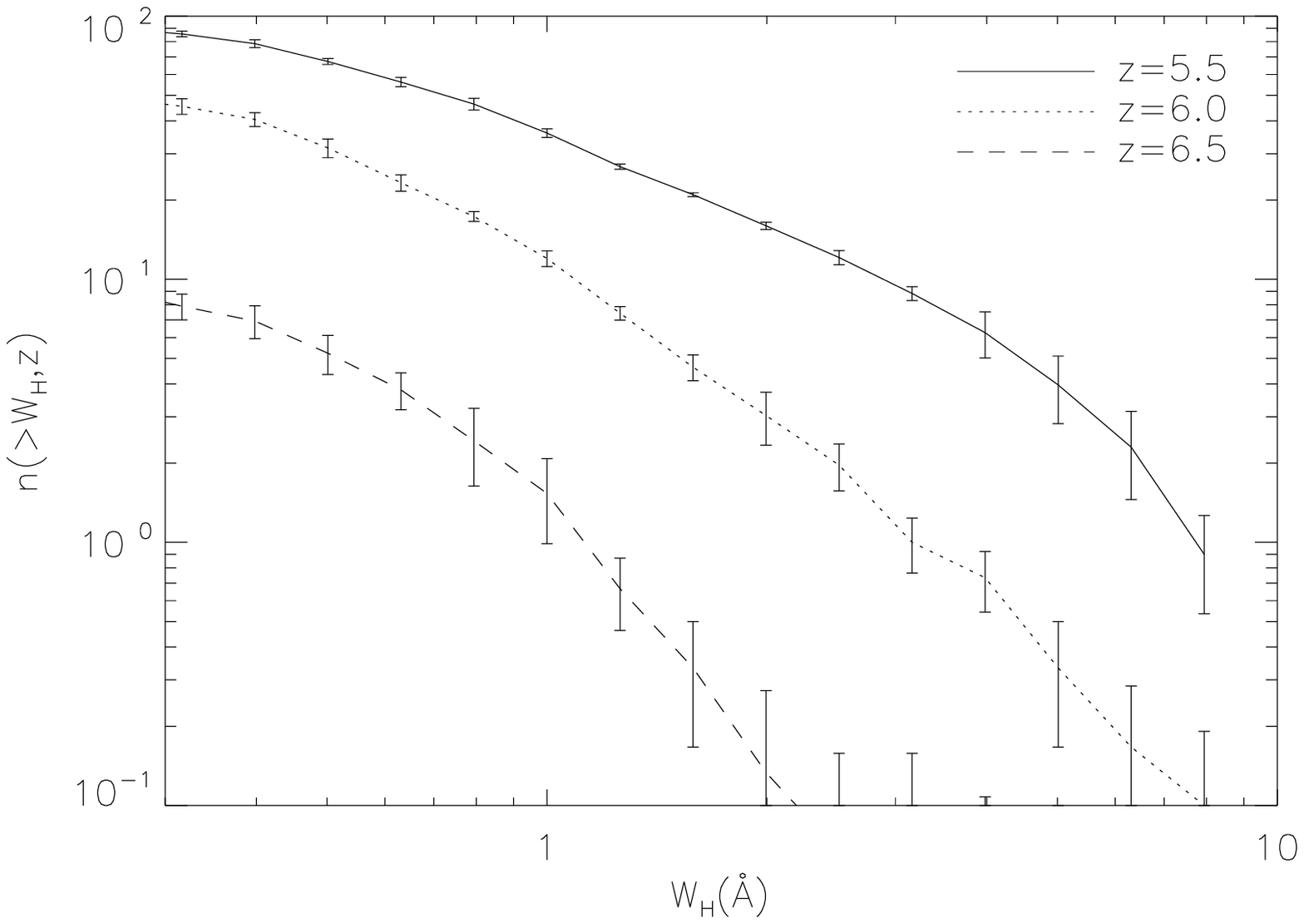,width=8.7truecm}} \caption{Number densities
$n(W,z)$, $n(>W,z)$, $n(W_H,z)$, and $n(>W_H,z)$ of leaks at
redshift $z=5.5$, 6.0 and 6.5 from top to bottom. The error bars are
the 1-$\sigma$ range given by Jackknife estimation, in
which each subsample contains 20 lines of the absorption spectrum.
}\label{ldis}
\end{figure*}

\begin{figure*}
\centerline{\psfig{figure=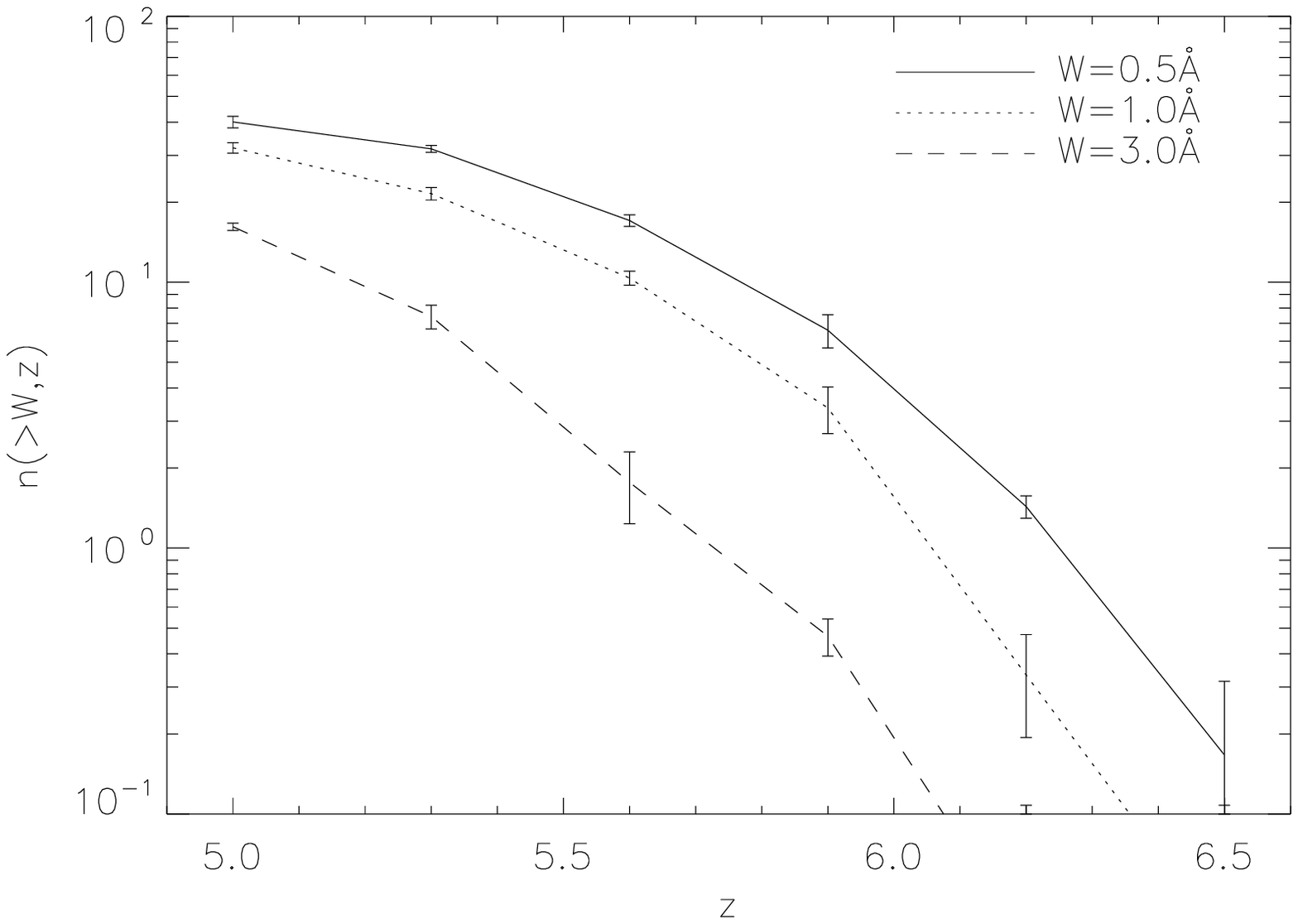,width=8.7truecm}
\psfig{figure=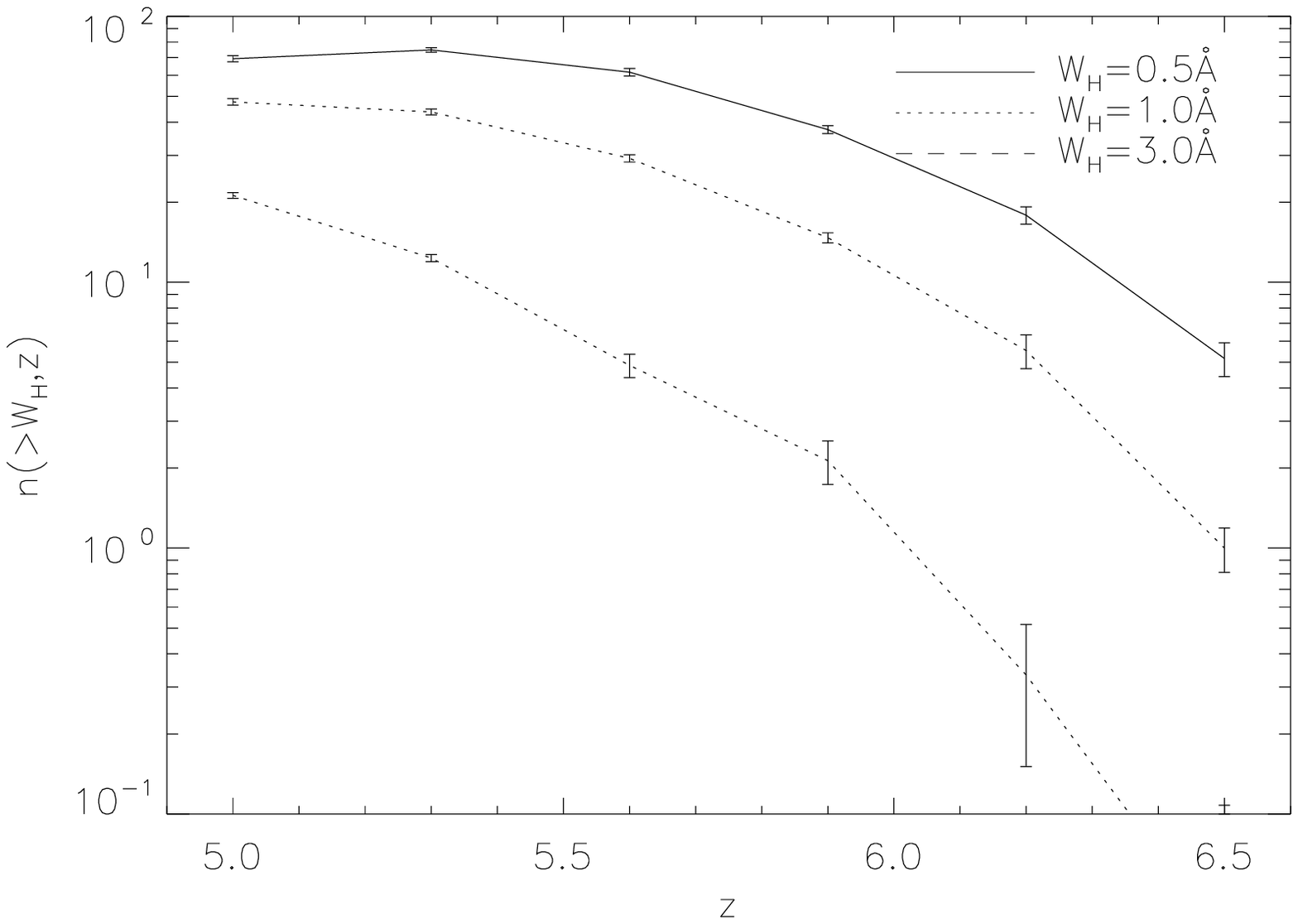,width=9.0truecm}}
 \caption{Redshift
evolution of number densities $n(>W,z)$, $n(>W,z)$, $n(W_H,z)$, and
$n(>W_H,z)$ of leaks with $W$, $\WH$ =0.5, 1.0, and 3.0 \AA. The
error bars are the 1-$\sigma$ range given by Jackknife 
estimation, in which each subsample contains 20 lines of the
absorption spectrum.}\label{zdis}
\end{figure*}

Similar to Ly$\alpha$ forests, we can define the cumulative number
densities $n(>W,z)$ and $n(>W_{\rm H},z)$ of leaks as the number of
leaks with widths larger than a given $W$ and $W_{\rm H}$ at $z$ per
unit $z$, respectively. The differential number densities are
$n(W,z)=dn(>W,z)/dW$ and $n(W_{\rm H},z)=dn(>W_{\rm H},z)/dW_{\rm
H}$. It should be pointed out that statistics of $W$ and $W_H$ are
not the same as the largest peak width statistics proposed by
Gallerani et al. (2007), which considered only the largest peak
width. A peak may contains more than one leaks, i.e. leak statistics
describe the details of the leaking area. The mean transmitted flux
at $z$ within $dz$ is $\bar{F}=\int_0^\infty n(W,z)WdW$. We
calculate the number densities of leaks in redshift range $z=5-6.5$.
In each redshift region we produce 100 light-of-sight samples to calculate
the density functions. The results are shown in Figure 5.
The errors are estimated by Jackknife method, i.e., the variance
over 5 subsamples, each of which contains 20 light-of-sight samples.

The number density $n(W,z)$ shown in the up-left panel of Figure 5
are similar to a Schechter function: they follow a power law at 
small $W$ and have a cut-off at large $W$.
The distribution of $n(\WH,z)$ at $\WH < 0.5$\A declines with 
decreasing $\WH$, this is because of the Jeans length smoothing.

The slope of $n(W,z)$  and $n(\WH,z)$ are smaller for small redshift.
It means the lack of low density voids with small size. 
That is, the increase of voids of small size is
less than voids of large size. This trend can also be seen from the
flattening of the cumulative density distributions $n(>\WH,z)$ and
$n(>W,z)$.

The redshift-evolution of the number densities $n(>W,z)$ and
$n(>W_H,z)$ of leaks for $W$, $W_{\rm H}$ =0.5, 1.0, and 3.0 \A are
shown in Figure \ref{zdis}. As has been seen in Figure \ref{ldis},
the number densities dramatically decrease at higher redshifts. The
evolution is more rapidly for large leaks: the number density
$n(>W=0.5$\AA,$\ z)$ drops by a factor of $\sim$ 5 when redshift
increasing from 5 to 5.8, while $n(>W=3$\AA,$ \ z)$ drops by a
factor of $\sim$ 60. The evolution trend is the same for the number
density $n(\WH,z)$. From the error bars of Figures \ref{ldis} and
\ref{zdis} we see that the predicted features of leaks would be able
used to compare with observed data set containing 20 or more
light-of-sight samples with the similar quality as simulation.

\subsection{Effects of Resolution and Noise }

\begin{figure*}
\centerline{\psfig{figure=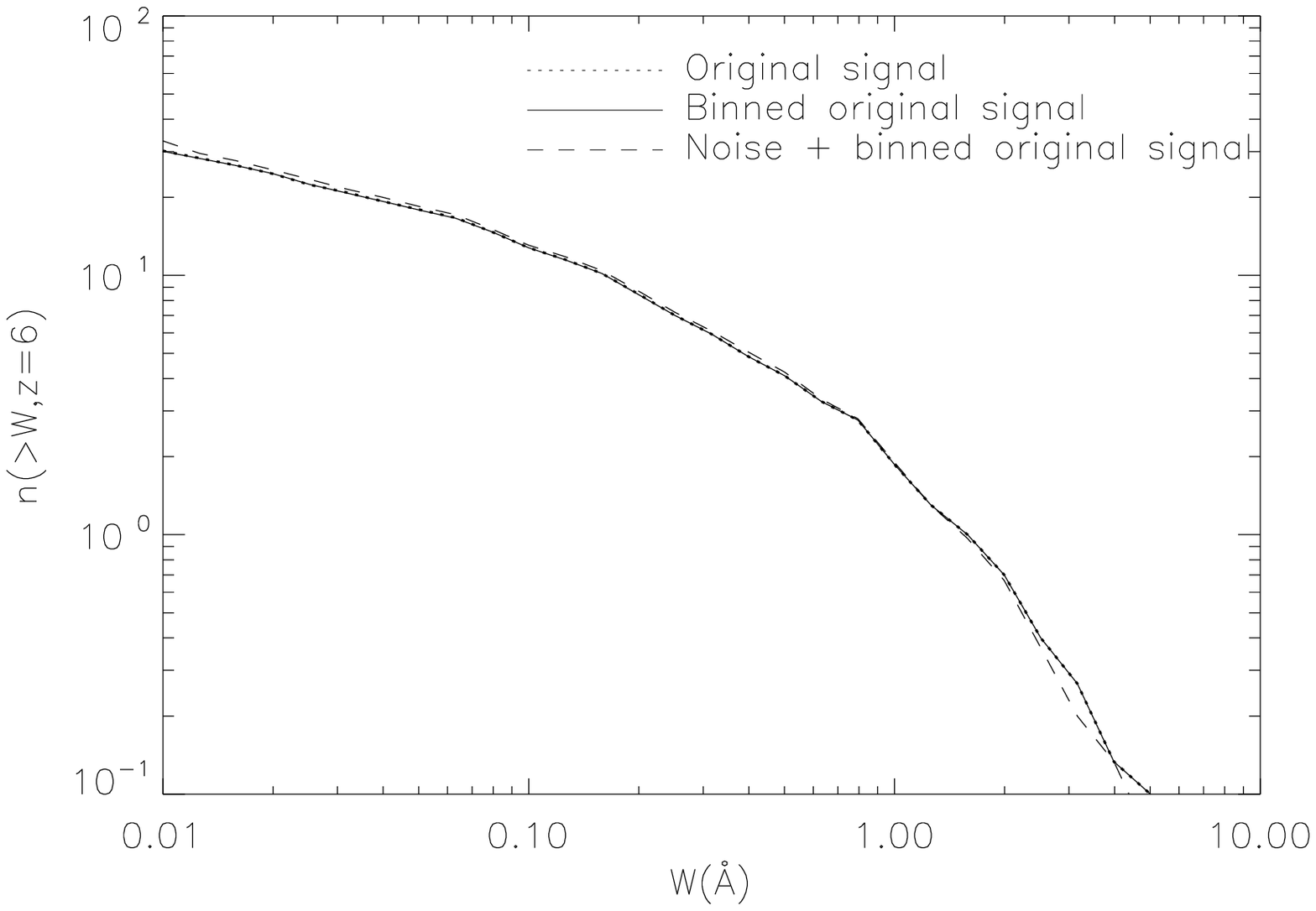,width=8.7truecm}
\psfig{figure=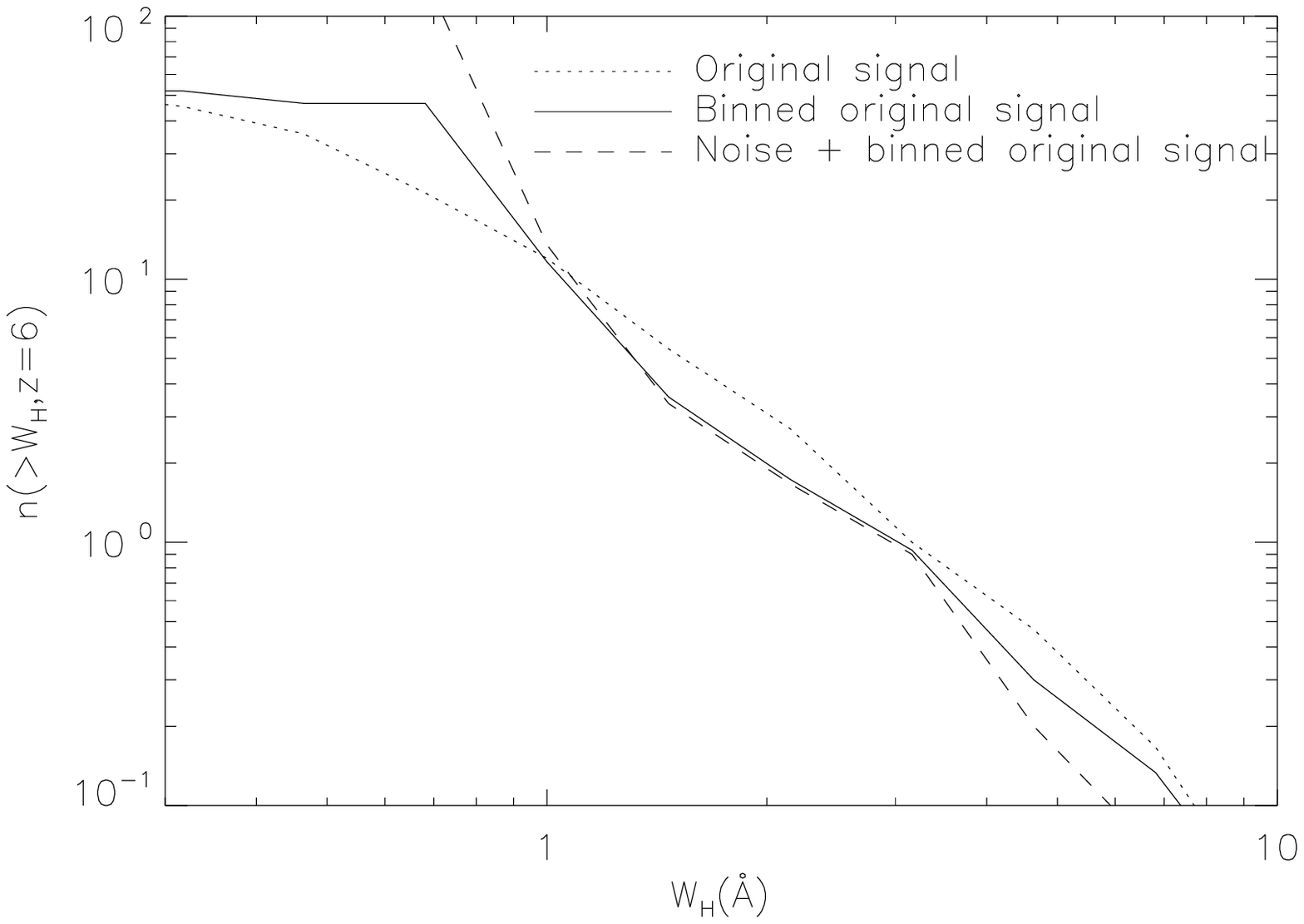,width=8.7truecm}} \caption{Effects of
resolution and noise on number densities $n(>W,z)$ (left panel) and
$n(>W_H,z)$ (right panel) for leaks at $z=6$.}\label{noise}
\end{figure*}

\begin{figure*}
\centerline{\psfig{figure=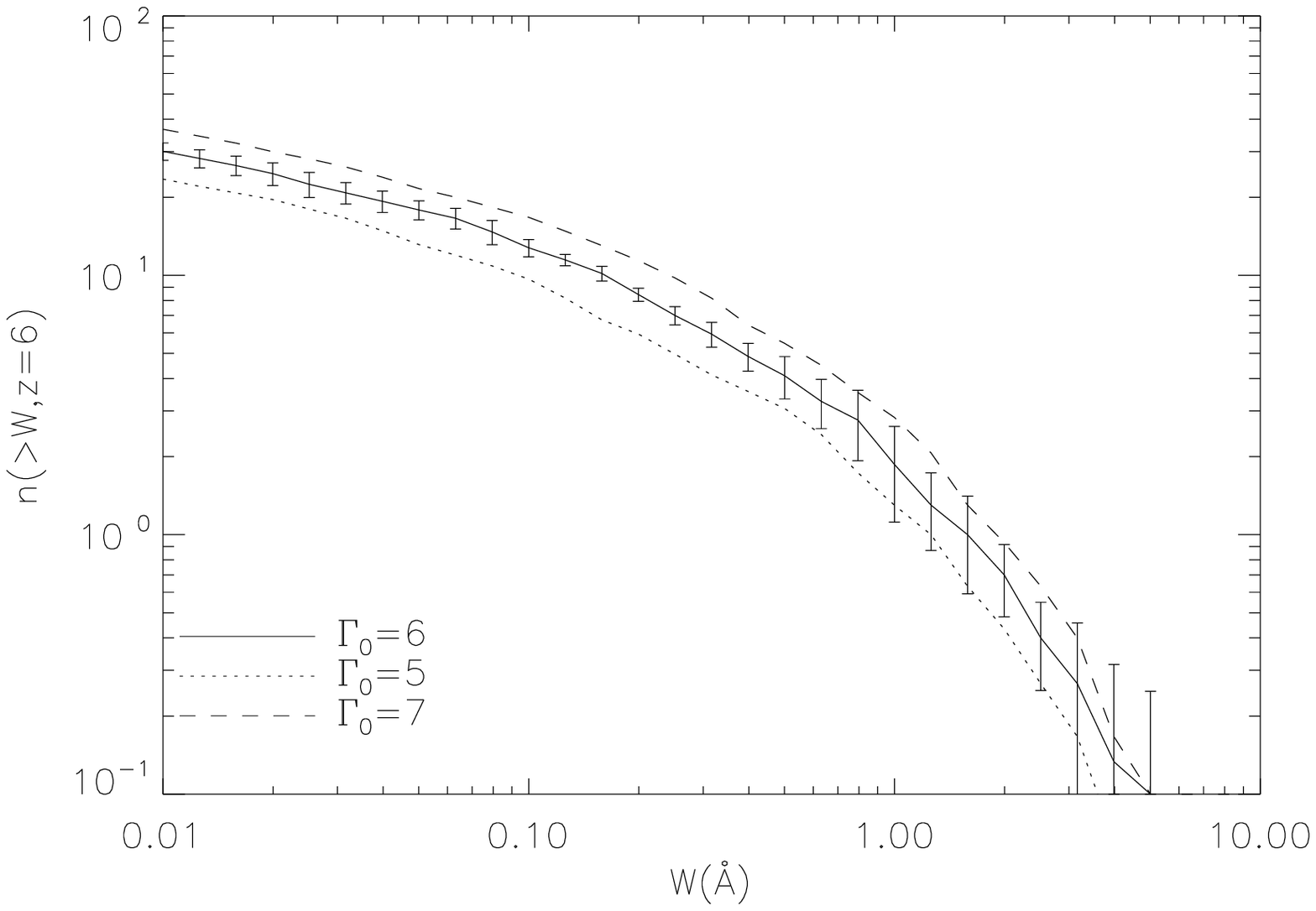,width=8.7truecm}
\psfig{figure=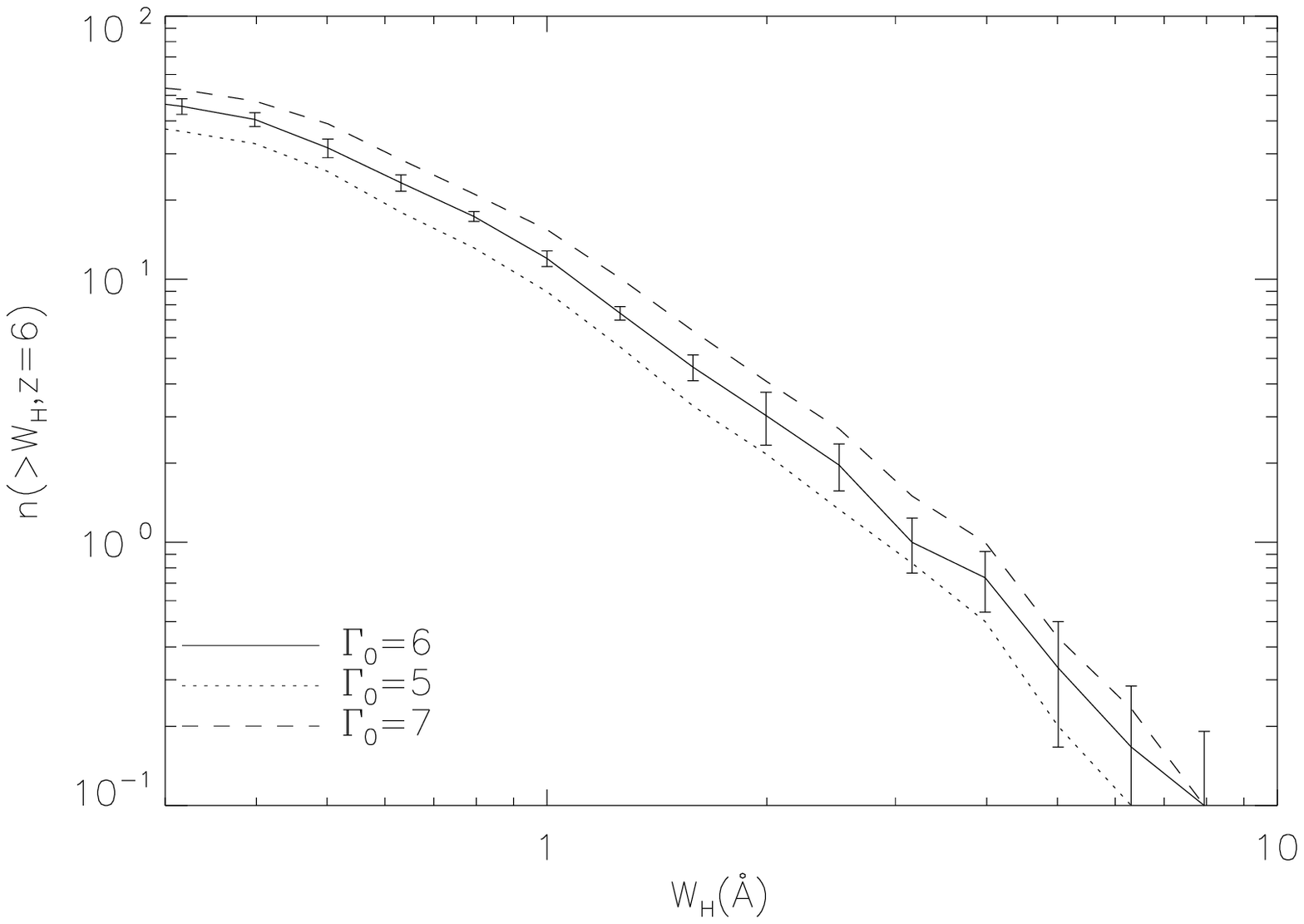,width=8.7truecm}} \caption{Number densities
$n(>W,z)$, and $n(>W_H,z)$ for samples at $z=6$ with $\Gamma_0=5$,
6, 7 $\times 10^{-12}$ s$^{-1}$, respectively. The error bars are
the same as Figure 5.}\label{gamma}
\end{figure*}

In this section, we study the observational and instrumental effects
on the statistics of leaks. Since both $W$ and $W_{\rm H}$ are
defined through the area under the profile of leaks, the effects of
resolution and noise would be small. To simulate the observational
effects, we bin the original data to a coarse grid corresponding to
a resolution of 20000, and we add Gaussian noises with
signal-to-noise ratio S/N=3 on binned pixels. The number densities
of leaks for the noisy binned samples are shown in Figures
\ref{noise}.

The effect of binning and noise is very small for $W$: the original
plot of $W$ actually is the same as the plot of binned $W$, and the
distribution of noisy $W$ is affected only when $W<0.01$\AA. The
effect of binning for $\WH$ is also small on scales larger than the
binning length. As expected, the noise effects for $\WH$ are
significant for $W_{\rm H}<0.4$ \AA. The noise effects for $\WH$ are
even smaller if we smooth the noisy sample. This is very different
from the PDF of the flux and the size of dark gaps, both of which
are heavily contaminated by instrumental resolution and
observational noises.

One can compare the uncertainty of $\Gamma_0$ with the effect of
noises. Figures \ref{gamma} shows the number densities for the UV
background amplitude $\Gamma_0$ = 5, 6, and 7 (eq.2), which
represent the allowed range of $\Gamma_0$. Different from noises,
the difference of $\Gamma_0$ will cause uncertainty in the whole
ranges of $W$ and $W_{\rm H}$. The uncertainties of number
densities are within a factor of 2 when the amplitude $\Gamma_0$
changes from 7 to 5. These uncertainties essentially are from the
mass density perturbations with long wavelengths (Paper I). The
error bars from the scattering of light-of-sight samples is also
shown in Figure \ref{gamma}. Therefore, the scattering of $\geq 20$
light-of-sight samples actually is less than the uncertainty of
$\Gamma_0$.

\section{\Lya LEAKS OF INHOMOGENEOUS IONIZING BACKGROUND}

In the early stage of reionization, ionizing photons are mainly in
ionized patches around UV ionizing sources, and therefore, the
spatial distribution of ionizing background is highly inhomogeneous
and has patchy structures. When the ionized patches spread over the
whole space, the ionizing background become uniform or
quasi-uniform, and so the ionizing background underwent an
inhomogeneous-to-uniform transition during the reionization. In this
section we study the effects of inhomogeneous ionizing background on
\Lya leak statistics.

\subsection{Models of Inhomogeneous Ionizing Background}

The first question is how to model the inhomogeneous spatial
distribution of the ionizing photon field. To our problem, the most
important property is the relation between the fields of mass
density and ionizing photon background or photoionization rate. If
the spatial fluctuation of photoionization rate $\Gxz$ is
statistically uncorrelated with the density field $\rho({\bf x},z)$,
the reionization of IGM will statistically be the same as a uniform
ionizing background, regardless of the details of the fluctuation of
$\Gxz$. In this case, the only effect of the fluctuations would be
to yield a larger variance in relevant statistics. However, as shown
in last sections, the uncertainty of the leak statistics is already
large even when the ionizing background is uniform, and therefore,
one would not be able to distinguish the fluctuating photon field
from inhomogeneous density field if they are uncorrelated.

Although many simulations on the UV photon field at the epoch of
reionization have been done, there still lack the results of the
correlation between photon and density fields. In this context, we
will consider the following four models on the statistical relation
between the inhomogeneous fields of photon and density, which are
mainly based on physical consideration of different stage of
reionization.

{\it Model 0}. The photoionization rate is spatially uniform. It
corresponds to the post-overlapping stage of reionization. This
model has been used in last two sections.

{\it Model I}. The photoionization rate at a give redshift is
assumed to be proportional to the density field of IGM, $\Gamma_{\rm
HI}=\Gamma_{\rm I} \rho $, $\Gamma_{\rm I}$ being a constant. This
model is motivated by the so-called inside-out scenario: high
density regions around UV sources are ionized first, and is most
probable at the early stage of reionization.

{\it Model II}. Just opposite to model I, the photoionization rate
at a give redshift is assumed to be inversely proportional to the
density field of IGM, $\Gamma_{\rm HI}=\Gamma_{\rm II} \rho^{-1}$,
$\Gamma_{\rm II}$ being a constant. This model comes from the
so-called outside-in scenario: under-dense regions are ionized first
(Miralda-Escude et al. 2000), which is applicable at the late stage
of reionization (Furlanetto \& Oh 2005).

In order to fit the observed effective optical depth $\tau_{eff}$ at
redshift $z=6$, we take the following parameters: $\Gamma_{\rm
I}=\Gamma_0\times3.53$, $\Gamma_{\rm II}=\Gamma_0/2.77$ where
$\Gamma_0$ is the photoionization rate at $z=6$ for uniform ionizing
background [eq.(2)].

{\it Model III  or patch model}. This model corresponds to the stage
before the overlapping of ionized patches. The IGM is almost fully
neutral except for some isolated ionizing patches. We model the
ionized patch as a Stromgren sphere: the neutral hydrogen fraction
is equal to 0 within the patch and equal to 1 outside the patch. The
mean radius of the Stromgren sphere is assumed to be $R_s\simeq 1.8$
comoving Mpc, which corresponding to an UV photon source with
luminosity $L= 5\times 10^{43}$ erg s$^{-1}$ with active time $10^7$
year and a $\nu^{-3}$ spectrum. To fit the observed optical depth,
we found there should be 2 patches every simulated box ($\Delta
z=0.3$). This number is actually consistent with the following
estimation
\begin{equation}
\frac{dN}{dz}= \frac{\pi R_s^2\phi(L)c}{H(z)},
\end{equation}
where $\phi(L)$ the comoving luminosity function, i.e. the 3-D number density
of sources with luminosity $L$. Using $\phi(5\times 10^{43})\sim 1.5
\times 10^{-3}$ Mpc$^{-3}$ (Bouwens et al. 2006), we have
$dN/dz\simeq 7$ and $(dN/dz)\Delta z\simeq 2$.

\subsection{\Lya Leaks of Inhomogeneous Ionizing Background}

\begin{figure*}
\centerline{\psfig{figure=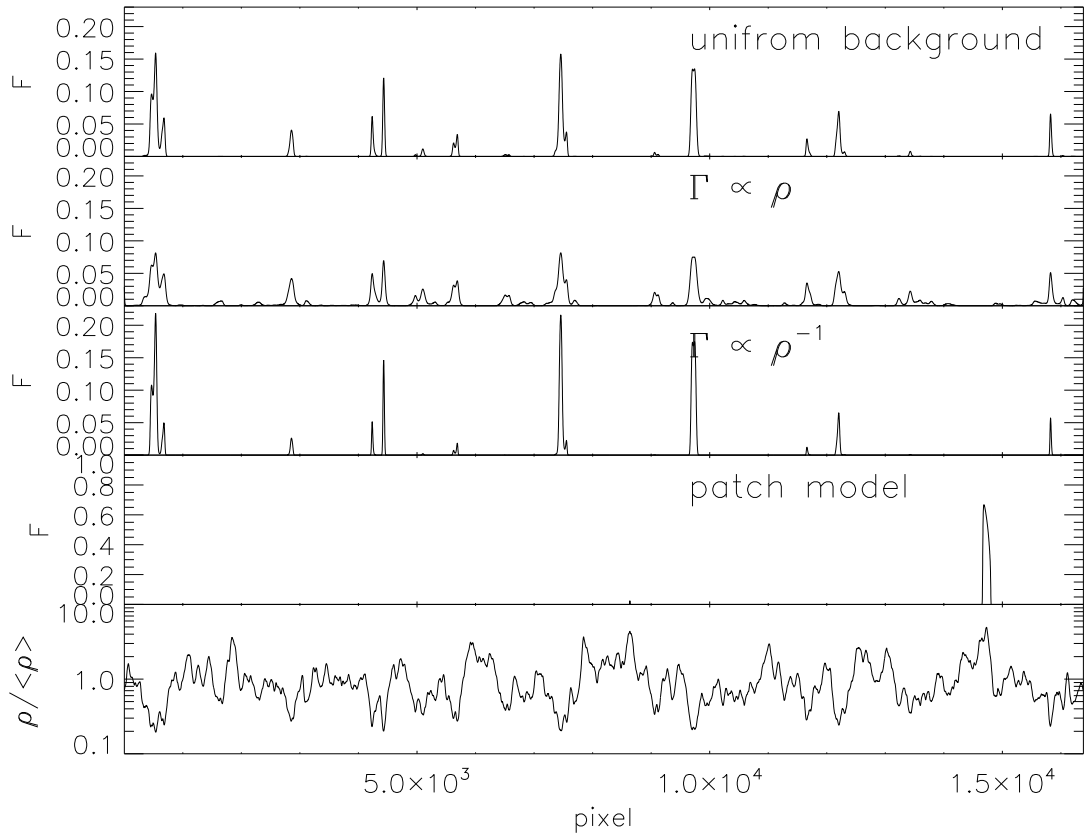,width=\hdsize}}
\caption{Transmitted flux $F$ given by uniform ionizing background
(top panel), the Model I (second panel), the Model II (third panel)
and the Model III (fourth panel). The density field $\rho$ of baryon
gas (bottom panel) is the same as Figure 2.}\label{s2}
\end{figure*}

For the four models described in the last section, we recalculate
the transmitted flux with the same underlying density and velocity
fields of fig2. The results are shown in Figure 9. As mentioned above, the
mean of transmitted flux for all models takes approximately the same
value of 0.004, which corresponds to an effective optical depth
$\simeq 5.5$. However, Figure 9 shows clearly different behaviors of
\Lya leaks in various models.

For models I and II, the leaks appear exactly at the same positions
as model 0 except some small leaks, which are more prominent for
model I. In other words, all the leaks of models 0, I and II are
from lowest density voids. Therefore, the three models have the same
distribution of dark gaps, and it is impossible to discriminate
among these models with the dark gap statistics.

However, the profiles of the leaks of model 0, I and II are
statistically different from each other.  For model I, the leaks
generally have larger width and lower height than model 0, while for
model II, the width of leaks generally is narrower than model 0, and
the height of leaks is larger than model 0. The reason is
straightforward. Comparing with model 0, model I gives a higher
$f_{\rm HI}$ at low density and lower $f_{\rm HI}$ at high density.
This leads to lower amplitude and broader width. For model II,  the
effect is just opposite to model I and yields  higher amplitude and
narrower width.

In the patch model, all the leaks come from ionized patches, within which ionizing sources are enclosed, and so the internal information of ionized
patches can be inferred from the leak statistics. The size of dark gaps is
actually given by the distance
between ionized patches. Generally,  \Lya leaks in the  patch model 
have a maximum flux $\simeq 0.6$, which is higher than the maximum
flux of other models, 0.2. This is because the neutral fraction
within the Stromgren sphere is much less that than other models.
This behavior is similar to the so-called proximity effect of QSOs
at low redshift (e.g., Rauch 1998). Thus, the statistics of the
maximum flux of \Lya leaks may be used to reveal the patchy origin
of \Lya leaks.

As we sample the size of patches along the sight of light according to the 
impact probability, the size of patches is smaller than $R_s$.
If the size is too small, the ionized patch will be opaque to 
\Lya photons due to the damping wing of the surrounding neutral 
hydrogen absorption (Miralda-Escude 1998). Obviously, it explains why only one 
leak is apparent in the patch model as displayed in Figure 9.

\subsection{Number Densities of \Lya Leaks of
Inhomogeneous Ionizing Background}

\begin{figure*}
\centerline{\psfig{figure=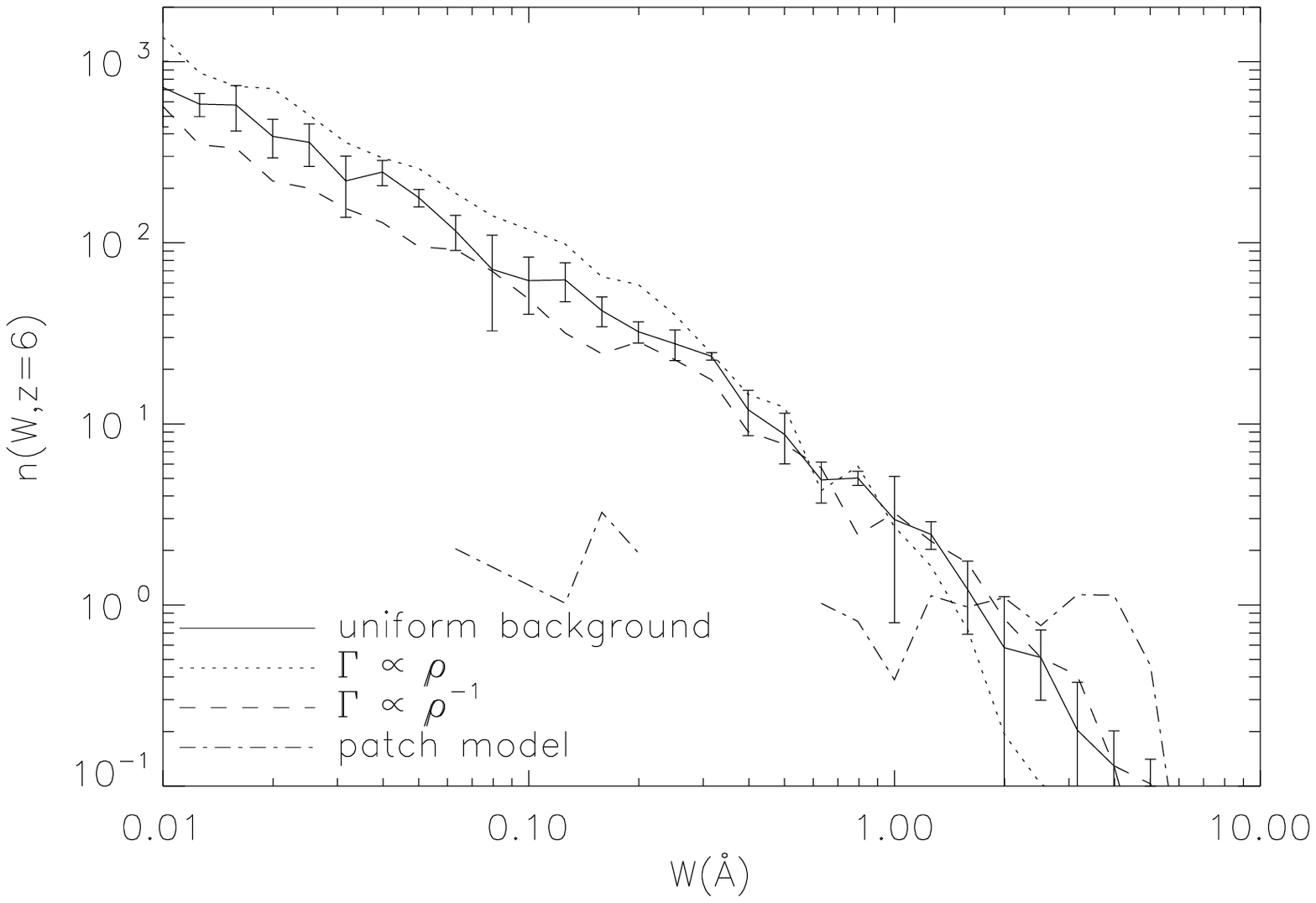,width=8.7truecm}
\psfig{figure=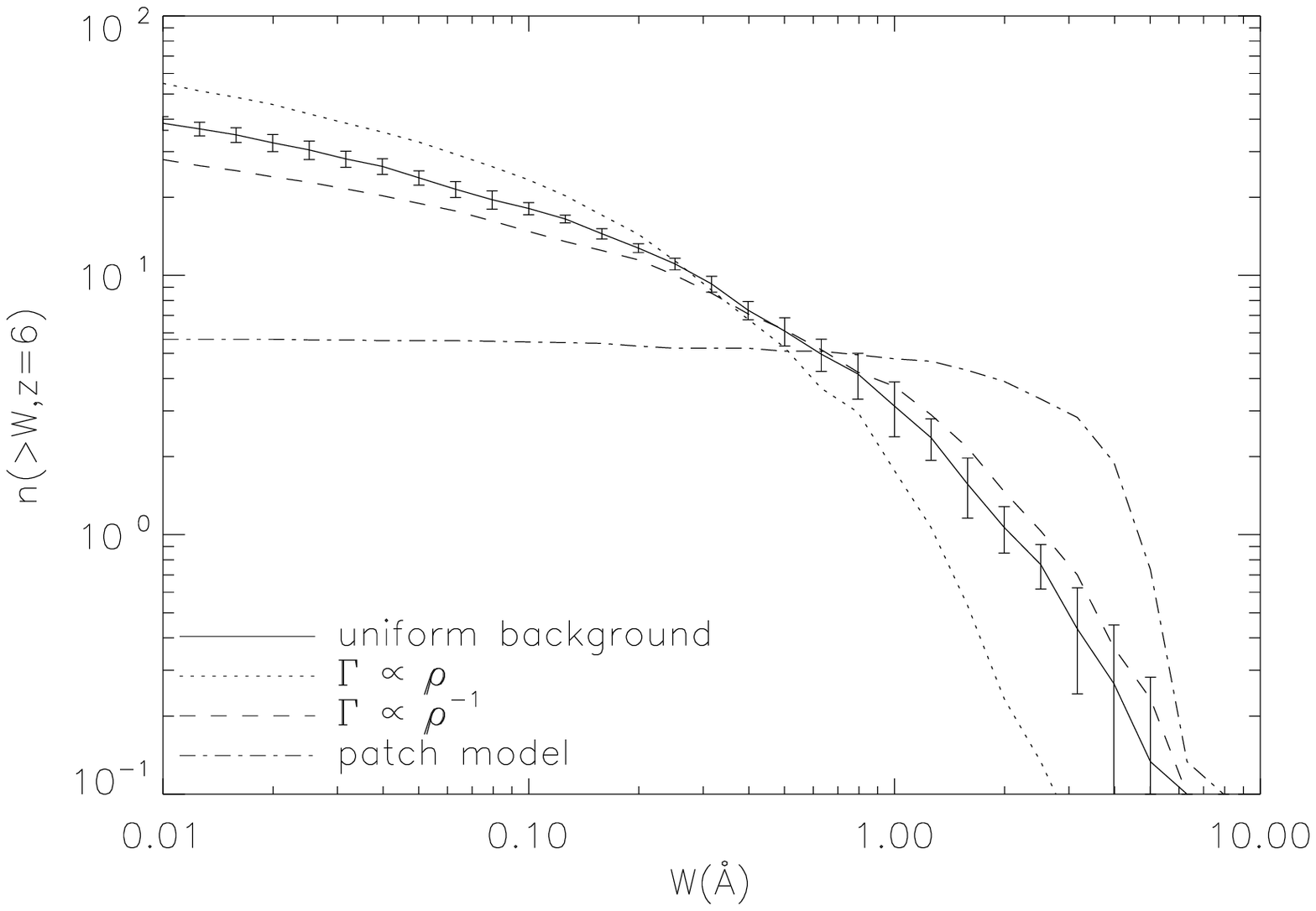,width=8.7truecm}}
\centerline{\psfig{figure=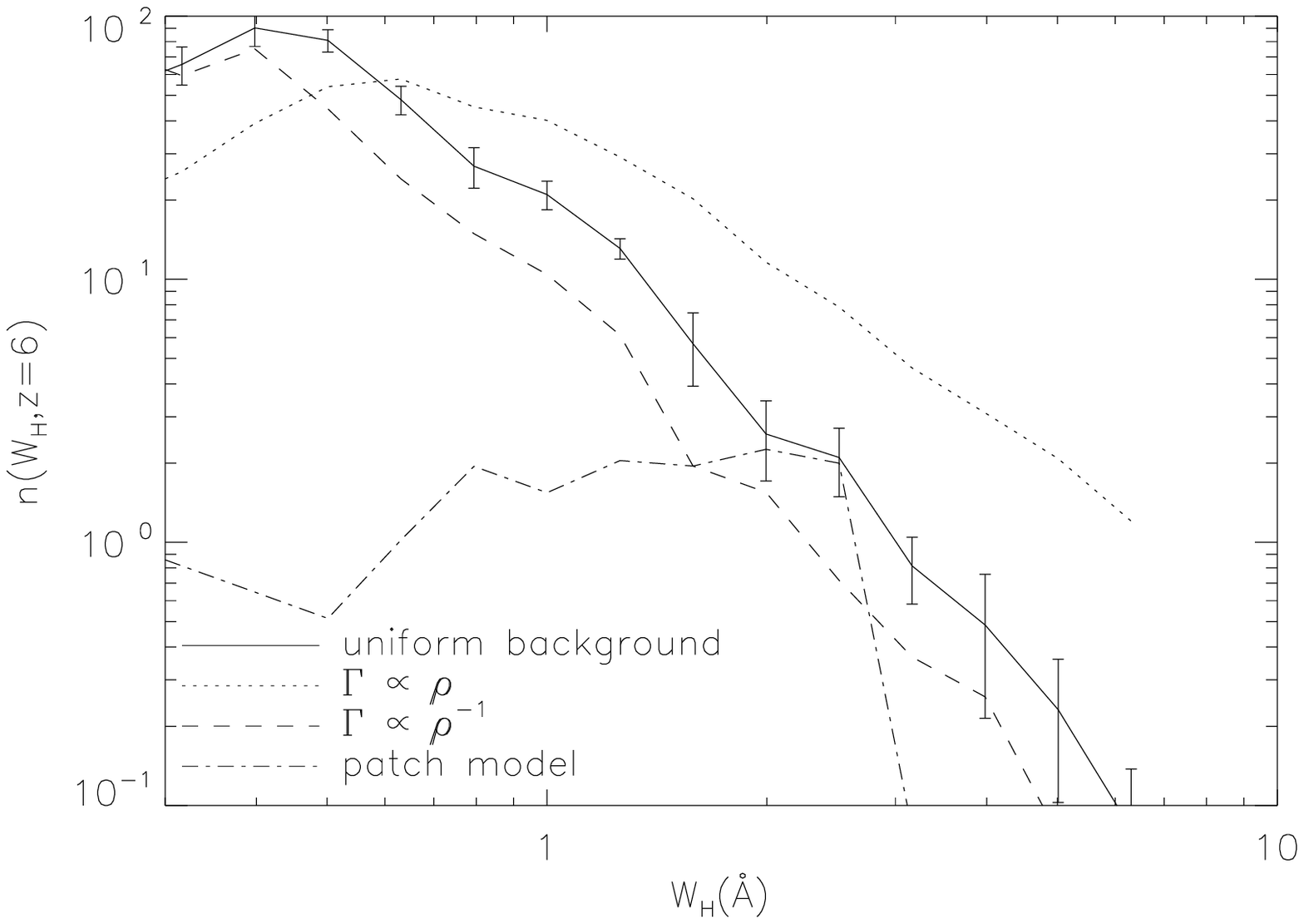,width=8.7truecm}
\psfig{figure=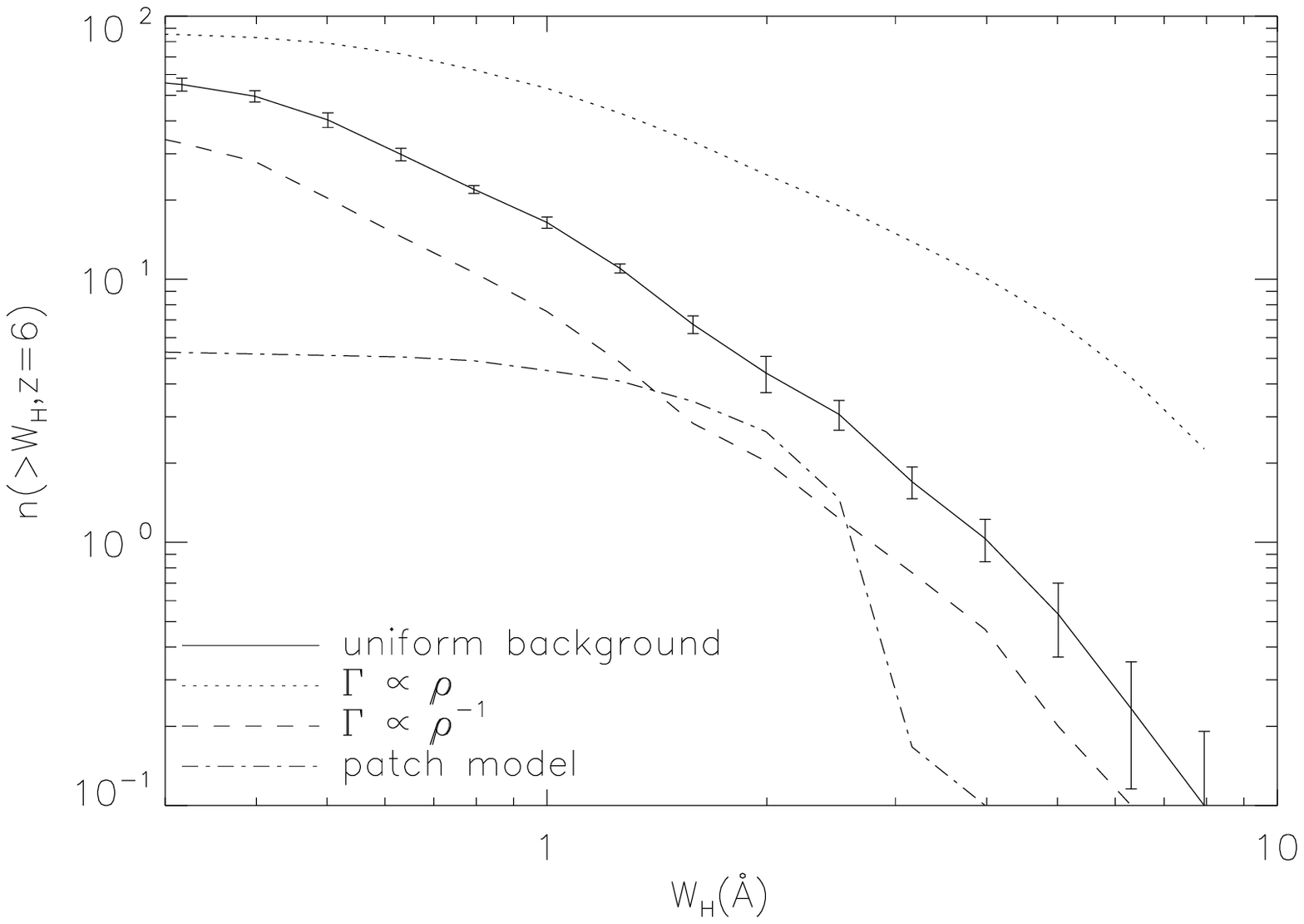,width=8.7truecm}} \caption{Number densities
$n(W,z)$,$n(>W,z)$, $n(W_H,z)$, and $n(>W_H,z)$ of leaks for three
fluctuating ionizing background models at $z=6$. The error bars of
the uniform background model are taken from Figure 5.}
\end{figure*}

We now calculate the number densities $n(W, z)$, $n(>W,z)$,
$n(W_{\rm H},z)$ and $n(>W_{\rm H}, z)$ for three inhomogeneous
ionizing background models at $z=6$, and the results are shown in
Figures 10. We can see from Figure 10 that the effects of different
models on the number densities of $W$ and $\WH$ are different. That
is, although all the models of 0, I, II and III give the same mean
effective optical depth, their leak distributions are different. For
clarifying, in Figures 10 we show only error bars for the curves of
model 0. It would be enough to show that these results can be tested
with data set containing about 20 light-of-sight data.

First, for model I, both of the number densities $n(W,z)$ and
$n(>W,z)$ have only small deviations from model 0. However,
$n(W_{\rm H},z)$ are significantly different from their counterparts
of model 0. The number density of leaks with $W_{\rm H}>1$\A is much
more than model 0. The basic feature of model I is to increase the
number of leaks with large $W_{\rm H}$ as shown in Figure 9.

Second, for model II, we see once again that the number densities
$n(W,z)$ and $n(>W,z)$ have only small deviations from model 0. Yet,
the number densities $n(\WH,z)$ and $n(>W_{\rm H},z)$ of the model
II are systematically lower than model 0. Therefore, the basic
feature of model II is to keep the total area under the profile of
leaks almost unchanged, but the widths of leaks are significantly
narrowed.

Finally, the behavior of patch model is very different from models
0, I and II. As expected, the patch model yields more leaks with
large $W$, and $n(W,z)$ shows a bump around $W=3-5$~\AA, which
characterize the area of the ionized patches. On the other hand, the
number densities of $n(\WH,z)$ and $n(>W_{\rm H},z)$ are lower than
model 0 because the characteristic size of ionized patches is less
than that of low density voids. It is interesting to note that for
the patch model, the tails of $n(W,z)$, $n(>W,z)$ and $n(W_{\rm H},
z)$, $n(>W_{\rm H}, z)$ are quite different from each other.
Generally, the tails of $n(W,z)$ and $n(>W,z)$ can extend to as large
as $W\simeq 5$\AA, while for $n(W_{\rm H},z)$ and $n(>W_{\rm H},z)$, there are 
no tails higher than 3 \AA. It results partially from 
the damping wing effect (Miralda-Escude 1998).

In summary, the statistical properties of leaks with respective to
$W$ and $\WH$ are sensitive to the details of ionizing photon field.
Combining the distribution of $W$ and $\WH$, the \Lya leaks would be
able to probe the origin of themselves, and thus reveal the
ionization state of IGM, the inhomogeneity of ionizing background,
and the evolution stage of reionization.

It should be pointed out that we considered only the patches of the
HI regions around high redshift galax
ies. The HI regions around
quasars or the proximity effect would also be the patches of
leaking. The mean luminosity of quasars probably is higher than
galaxies, and therefore, the above-mentioned feature of $n(\WH,z)$
and $n(>W_{\rm H},z)$ would be more prominent for the patches of
quasars. The damped Ly$\alpha$ absorption system is important for
modeling low redshift Ly$\alpha$ absorption. Since these systems
have high column density of neutral hydrogen, it will not contribute
to leaking either in density void models or patch model.

\section{DISCUSSION AND CONCLUSION}

We show that the Ly$\alpha$ absorption spectra of UV photon emitters
at redshifts around $z\sim6$ can be decomposed into \Lya leaks,
which come from either lowest density voids or ionized patches
containing ionizing sources. The \Lya leaks are well defined and
described by the equivalent width $W$ and the width of half area
$\WH$. Since both $W$ and $\WH$ are defined through integrated
quantities, the distributions of \Lya leaks in terms of $W$ and
$\WH$ are stable with respect to observational noises and
instrumental resolution. Although the number densities $n(W,z)$,
$n(>W,z)$, $n(W_{\rm H},z)$, and $n(>W_{\rm H},z)$ evolve very
rapidly at redshift $z\simeq 6$, these statistics are measurable up
to $z=6.5$.

If the \Lya leaks come from lowest density voids, the IGM should be
still highly ionized and the ionizing background is almost uniform;
in contrast, if the leaks come from isolated ionized patches, the
ionizing background should be inhomogeneous, and the reionization is
still in the overlapping stage. Therefore, the origin of \Lya leaks
is crucial to understand the history of reionization.

We show that the statistics with $W$ and $\WH$ are sensitive to the
origin of \Lya leaks, because the \Lya leaks are sensitive to the
correlation between photon and density fields. Based on physical
consideration of reionization, we studied four phenomenological
models of the photon field: the uniform ionizing background (model
0), the photoionization rate $\GHI$ is proportional to the density
$\rho$ (model I), $\GHI$ and $\rho$ are anti-correlated (mode II),
and ionized regions only given by Stromgren sphere around ionizing
photo sources (patch model). We found that, although all the four
models can fit the observed mean and variance of optical depth at
$z\simeq 6$, the width distribution of \Lya leaks show very
different behaviors.

For model 0, I, and II, most of \Lya leaks are from lowest density
voids, and the distribution of dark gaps are similar. However, the
properties of individual \Lya leaks are different. Model I gives
broader width $\WH$ than model 0; model II gives narrower width
$\WH$ than model 0. For patch model, the \Lya leaks are mainly from
ionized sphere, and they generally have higher maximum of
transmitted flux than other models. There is a bump in the
distribution of the equivalent width $W$, which characterize the
intensity of UV photons within ionized patches. \Lya leaks from
ionized patches will provide the information of ionized patches,
such as their intensity and size, and constrain the properties of UV
sources that contribute most of ionizing photons of the
reionization.

Finally, we point out that similar analysis is also applicable to
the reionization of HeII. The optical depth of HeII \Lya evolves
rapidly at $z\simeq2-3$. It reaches $\simeq 5$ at $z\simeq 3$. That
is, the evolution of HeII reionization at $z\simeq 3$ would be
similar to that of H at $z\simeq 6$. Observed sample of HeII \Lya
absorption indeed show structures similar to the leaks of hydrogen
absorption spectrum (Zheng et al. 2004). One can also draw the
information of background photons at the energy band of HeII
ionization.


\section*{ACKNOWLEDGEMENT }

This work is supported in part by the US NSF under the grant
AST-0507340. LLF acknowledges support from the National Science
Foundation of China (NSFC).


\label{lastpage}


\begin{thebibliography}{}

\bibitem[]{Bah00}
Bahcall N.A., Cen R., Dav\'e R., Ostriker J.P., Yu Q., 2000, \apj ,
541, 1

\bibitem[]{Bechtold} Bechtold, J. 1994, \apjs, 91, 1

\bibitem[]{Becker} Becker, G. D., Rauch, M. \& Sargent, W. L. W. 2006,
    astro-ph/0607633

\bibitem[]{Ber} Berera, A. \& Fang, L. Z. 1994, Phys. Rev. Lett., 72, 458


\bibitem[]{Bi1997} Bi, H.G. \& Davidsen, A. F. 1997, \apj, 479, 523.

\bibitem[]{Bi1995} Bi, H.G., Ge, J. \& Fang, L.Z. 1995, \apj, 452, 90

\bibitem[]{Bi2003} Bi, H.G., Fang, L.Z., Feng, L.L. \& Jing, Y.P. 2003,
    \apj, 598, 1

\bibitem[]{Bolton} Bolton, J. S., Haehnelt, M. G., Viel, M., Springel, V.
    2005, \mnras, 357, 1178

\bibitem[]{Bouwens} Bouwens, R. J., Illingworth, G. D., Blakeslee, J. P.,
    Franx, M. 2006, \apj, 653, 53

\bibitem[]{Choudhury2001} Choudhury, T.R., Srianand, R. \& Padmanabhan T., 2001,
\apj, 559, 29

\bibitem[]{Choudhury2005} Choudhury, T.R. \& Ferrara,
A. 2005, \mnras, 351, 577

\bibitem[]{Ciardi}  Ciardi, B., Stoehr, F. \& White, S. D. M. 2003,
\mnras,343,1101

\bibitem[]{Croft} Croft, R. A. C. et al. 2002, \apj, 581, 20

\bibitem[]{Ellis} Ellis, R. S. 2007, astro-ph/0701024, First Light in
Universe, Saas-Fee Advanced Course 36, Swiss Soc. Astrophys. Astron

\bibitem[]{Fan2002} Fan, X. et al. 2002, \aj, 123, 1247

\bibitem[]{Fan2006} Fan, X. et al. 2006, \aj, 132, 117

\bibitem[]{Furlanetto2004}  Furlanetto, S., Hernquist, L. \& Zaldarriaga, M. 2004,
    \mnras, 354, 675

\bibitem[]{Furlanetto2005}  Furlanetto, S. \& Oh, S. P. 2005, \mnras, 363, 1031

\bibitem[]{Gallerani} Gallerani, S., Choudhury, T.R. \& Ferrara, A. 2006,
    \mnras, 370, 1401

\bibitem[]{Gallerani2} Gallerani, S., Ferrara, A., Fan, X. \& Choudhury, T.R.
     2007, arXiv:0706.1053.

\bibitem[]{Gnedin} Gnedin, N. 2004 \apj, 610, 9

\bibitem[]{He2004} He, P., Feng, L.L. \& Fang, L.Z. 2004, \apj, 612, 14

\bibitem[]{He} He, P., Feng, L. L., Shu, C. W. \& Fang, L. Z. 2006,
Phys, Rev. Lett. 96, 051302

\bibitem[]{Haiman2002} Haiman, Z. 2002, \apj, 576, L1

\bibitem[]{Hui1997} Hui, L. \& , Gnedin, N.Y. 1997 \mnras, 292, 27

\bibitem[]{Jena2005} Jena, T. et al. 2005, \mnras, 361, 70

\bibitem[]{Jon} Jones, B.T. 1999, \mnras, 307, 376

\bibitem[]{Kohler2007} Kohler, K., Gnedin, N. Y., \& Hamilton, A. J. S. 2007,
    \apj, 657, 15

\bibitem[]{Lidz2006} Lidz, A., Oh, S. P. \& Furlanetto, R. 2006, \apj, 639,
L47

\bibitem[]{Liu2006} Liu, J., Bi, H., Feng, L.-L. \& Fang, L.-Z. 2006, \apj,
645, L1 (PaperI)

\bibitem[]{Liu2007} Liu, J.R. \& Fang, L.Z. 2007, arXiv:0707.2620,
\apj, in press

\bibitem[]{Mellema} Mellema, G., Iliev, I. T., Pen, U.-L., Shapiro, P. R.
2006, \mnras, 372, 679

\bibitem[]{Mir} Miralda-Escude, J. 1998, \apj, 501, 15

\bibitem[]{Miralda2000} Miralda-Escude, J.,  Haehnelt, M. \& Rees, M. J. 2000,
    \apj, 530, 1

\bibitem[]{Oh2005} Oh, S.P., \& Furlanetto, S.~R.\ 2005, \apj, 620, L9

\bibitem[]{Page} Page, L. 2007, \apjs, 170, 335.

\bibitem[]{Pan} Pando, J.,  Feng, L. L., Jamkhedkar, P.,
  Zheng, W., Kirkman, D., Tytler, D.  \& Fang, L. Z. 2002, \apj, 574, 575


\bibitem[]{Paschos2005} Paschos, P., \& Norman, M. L., 2005, ApJ, 631, 59

\bibitem[]{Qiu2006} Qiu, J.M., Shu, Q.W., Feng, L.L. \& Fang, L.Z. 2006,
    in preparation

\bibitem[]{Rauch1997} Rauch, M. et al. 1997, \apj, 489, 7

\bibitem[]{Rauch1998} Rauch, M. 1998, \araa, 36, 267

\bibitem[]{Razomov2002} Razoumov, A. O., Norman, M. L., Abel, T., \& Scott, D.,
2002, ApJ, 572, 695

\bibitem[]{Seljak2005} Seljak, U. et al. 2005, PhRvD, 71, 103515

\bibitem[]{Shapiro2006}  Shapiro, P. R., Iliev, I. T., Alvarez, M. A.,
    Scannapieco, E. 2006, \apj, 648, 922

\bibitem[]{Sokasian2003} Sokasian, A, Abel, T., Hernquist, L, Springel, V. 2003,
    \mnras, 344, 607S

\bibitem[]{Songalia2002} Songaila, A.  \& Cowie, L.  2002, \aj, 123, 2183

\bibitem[]{Viel2002} Viel, M., Matarrese, S., Mo H.J., Haehnelt, M.G.,
Theuns, T., 2002, \mnras, 329, 848

\bibitem[]{Viel2006} Viel, M., S., Haehnelt, M.G., Lewis, A. 2006, \mnras,
370, L51

\bibitem[]{Wyithe2005} Wyithe, J, S. B., \& Loeb, A.\ 2005, \apj, 625, 1

\bibitem[]{Yang}  Yang, X. H., Feng, L. L., Chu, Y. Q. \&
     Fang, L. Z. 2001, \apj, 560, 549

\bibitem[]{Zheng2004} Zheng, W. et al. 2004, \apj, 605, 631

\end{thebibliography}
\end{document}